\documentclass[namedreferences]{SolarPhysics}
\usepackage[optionalrh]{spr-sola-addons} 
\usepackage{graphicx}        
\usepackage{color}           
\usepackage{url}             

\begin{document}

\begin{article}

\begin{opening}

\title{Multiwavelength Study on Solar and Interplanetary Origins of the Strongest Geomagnetic
Storm of Solar Cycle 23}

\author{Pankaj~\surname{Kumar}$^{1}$\sep
        P.K.~\surname{Manoharan}$^{2}$\sep
        Wahab~\surname{Uddin}$^{3}$    
       }
\runningauthor{P. Kumar et al.}
\runningtitle{Origin of strongest geomagnetic storm of solar-cycle 23}
\institute{$^{1}$ Korea Astronomy and Space Science Institute (KASI),Hwaamdong, Yuseong-Gu,
 Daejeon-305-348, Republic of Korea, email: \url{pankaj@kasi.re.kr}}
\institute{$^{2}$ Radio Astronomy Centre, National Centre for Radio Astrophysics, Tata Institute of Fundamental Research, Udhagamandalam (Ooty)-643 001, India, email: \url{mano@ncra.tifr.res.in}}
\institute{$^{3}$Aryabhatta Research Institute of Observational Sciences (ARIES), Manora Peak, Nainital-263129, India}

\begin{abstract}
We study the solar sources of an intense geomagnetic storm of solar cycle 23 that occurred on 20 November 2003, based on ground- and space-based multiwavelength observations. The coronal mass ejections (CMEs) responsible for the above geomagnetic storm originated from the super-active region NOAA 10501. We investigate the H$\alpha$ observations of the flare events made with a 15 cm solar tower telescope at ARIES, Nainital, India. The propagation characteristics of the CMEs have been derived from the three-dimensional images of the solar wind ({\it i.e.}, density and speed) obtained from the interplanetary scintillation data, supplemented with other ground- and space-based measurements. The TRACE, SXI and H$\alpha$ observations revealed two successive ejections (of speeds $\approx$350 and $\approx$100 km s$^{-1}$),  originating from the same filament channel, which were associated with two high speed CMEs ($\approx$1223 and $\approx$1660 km s$^{-1}$, respectively). These two ejections generated propagating fast shock waves ({\it i.e.}, fast drifting type II radio bursts) in the corona. The interaction of these CMEs along the Sun-Earth line has led to the severity of the storm. According to our investigation, the interplanetary medium consisted of two merging magnetic clouds (MCs) that preserved their identity during their propagation. These magnetic clouds made the interplanetary magnetic field (IMF) southward for a long time, which reconnected with the geomagnetic field, resulting the super-storm ($Dst_{\rm peak}=-472$ nT) on the Earth.
\end{abstract}
\end{opening}

\section{Introduction}
 Solar flares are the transient explosions in the solar atmosphere during which magnetic energy stored in twisted and sheared fields is suddenly released in the form of electromagnetic radiation and particle acceleration.   Gigantic clouds of ionised gas and magnetic field ejected out from the corona in association with flares are known as coronal mass ejections (CMEs). For example, a CME in the interplanetary medium (known as ICME) is also referred to as the magnetic cloud (MC), which is the ejection of a flux rope into the solar wind. It is defined by relatively intense magnetic field, large and smooth rotation of field direction over an approximate size of 0.25 AU at 1 AU, and low proton temperature \cite{burlaga1981}. The intensity and the duration of the southward component of the interplanetary magnetic field (IMF) is the most critical factor to determine the geoeffectivity of MCs. The magnetic reconnection taking place at the boundary of the magnetosphere is controlled by the southward component of the CME-associated field. After the solar wind energy is pumped into the magnetosphere by magnetic reconnection, geomagnetic storms may take place due to the formation of a ring current circulating around the Earth above the equator. Several authors have reported that the southward component of IMF ($B_z<0$) and CME speed are the main driver to produce intense geomagnetic storms \cite{tsurutani1988}. However, the magnetic topology of a MC depends on its helicity, inclination of the axis with respect to the ecliptic plane,  and polarity, {\it i.e.} whether the field rotates from the south at the leading edge to the north at the trailing edge (SN-polarity) or vice verse (NS-polarity). The geoeffectiveness of MCs with different topologies is greatly different. But still it is challenging to predict the occurrence of intense storms in terms of solar and interplanetary parameters.  Moreover, at the sheaths behind the interplanetary shocks, the magnetic field can get amplified and lead to intense southward $B_z$ component. This region is equally important for creating magnetic storms. Many researchers ({\it e.g.}, \opencite{echer2008a}) have shown that it is about 50$\%$-50$\%$ in geoeffectiveness for storms of $Dst\le-100$ nT intensity. The sheath is not part of the ICME. It is plasma and fields that have been swept up by the passage of the ICME. Thus, CMEs are the integral part of space weather.

During October-November 2003, a series of solar eruptions occurred from three solar active regions ({\it i.e.}, NOAA AR 10484, 486, 488). AR 10484 returned as AR 10501 (located at $\approx$N00 E18) and produced the largest geomagnetic storm of solar cycle 23 on 20 November 2003. The flares/CMEs from this AR have been studied by several authors \cite{gopal2005,mostl2008,srivastava2009,chandra2010,kumar2010,sch2011}.  \inlinecite{gopal2005} suggested that the largest geomagnetic storm was caused by a fast and wide CME originating close to the disk center. It resulted in a highly inclined MC with its axial field pointing almost always southward, which reconnected with Earth's magnetosphere. \inlinecite{mostl2008} compared the properties of the source region with those inferred from satellite observations near the Earth (SOHO and TRACE data), with ground-based data. They modeled the near-Earth observations with the Grad-Shafranov reconstruction technique using a novel approach in which they optimized the results with two-spacecraft measurements of the solar wind plasma and magnetic field made by ACE and {\it Wind} and found that the magnetic cloud orientation at 1 AU is consistent with an encounter with the heliospheric current sheet. In a study on the flare site characteristics, \inlinecite{srivastava2009} estimated the temporal variation in magnetic flux, energy, and magnetic field gradient for the source active region using MDI magnetograms, which provided evidence that the flare associated with the CME occurred at a location marked by high magnetic field gradient and led to the release of free energy stored in the active region. 

In this study, we will investigate the flare events on 18 November 2003 using H$\alpha$ images associated with the on-disk signatures to infer the constituent of CMEs as well as to understand their propagation characteristics and  their consequences at 1 AU using ground- and space-based measurements.
 

\begin{figure}
\centering
\includegraphics[width=0.45\textwidth]{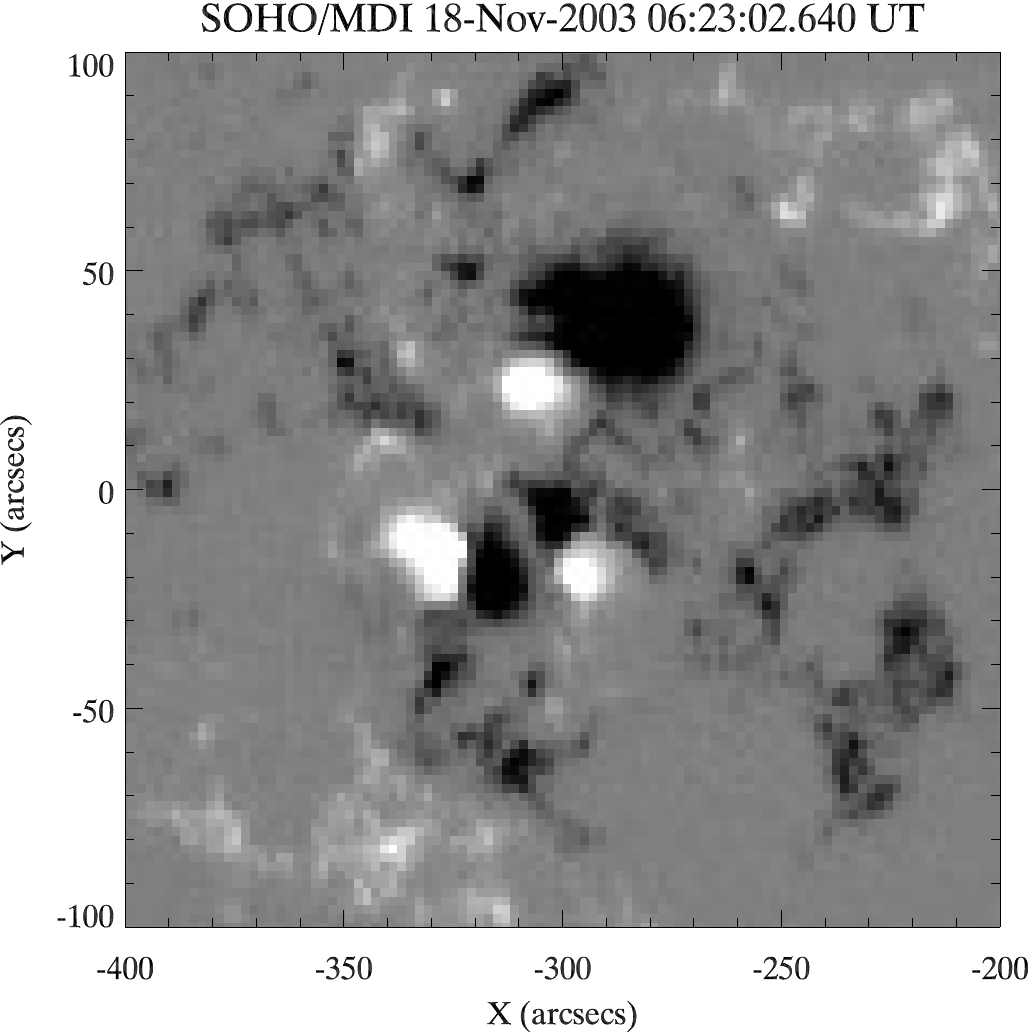}
\thicklines
$\color{red} \put(-83,73){\circle{200}}$
\includegraphics[width=0.45\textwidth]{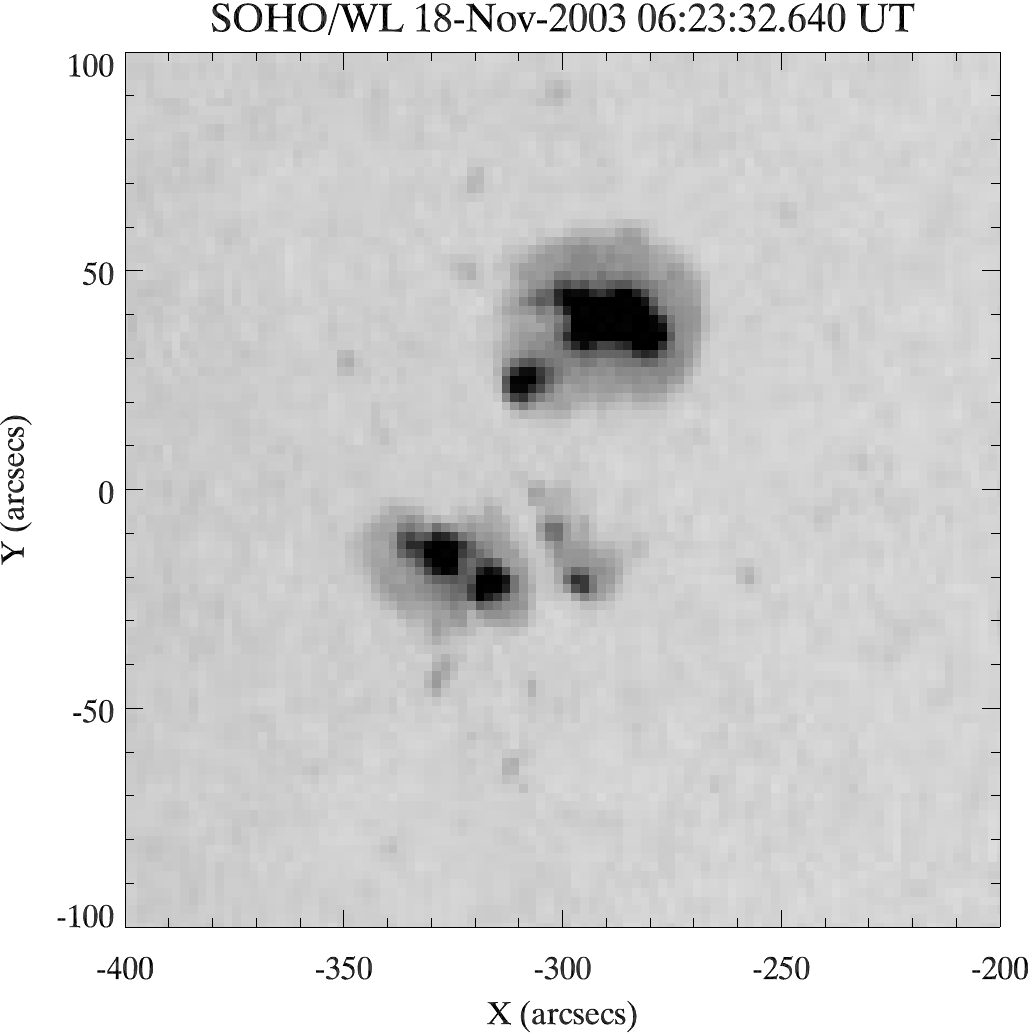}
$\color{red} \put(-83,73){\circle{200}}$
\caption{SOHO/MDI and white-light images of the active region NOAA 10501 on 18 November 2003. The quadrupolar magnetic field structure shown inside the circle was the source region of flares/CMEs that occurred in the active region.} 
\label{ar}
\end{figure}


\section{Observations and Data}
\begin{figure}
\centering
\includegraphics[width=0.8\textwidth]{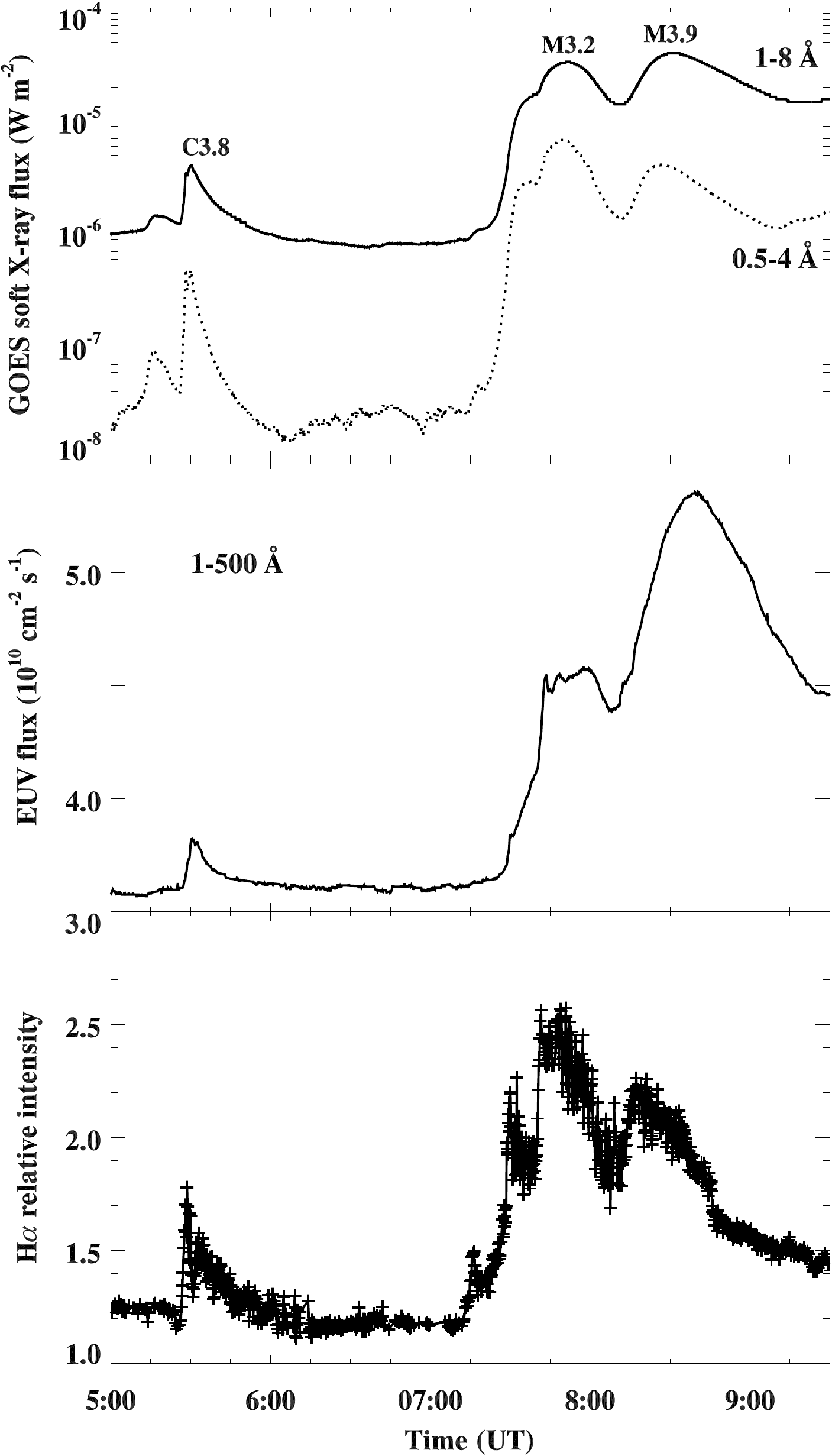}
\caption{GOES soft X-ray flux, EUV flux, and H$\alpha$ relative intensity profiles for the flares on 18 November 2003.}
\label{fluxes}
\end{figure}
Figure \ref{ar} shows the Michelson Doppler Imager (MDI) magnetogram and white-light image of the active region NOAA 10501 on 18 November 2003. As shown by the manetogram, the active region was rather complex with $\beta\gamma\delta$ magnetic configuration. The active region was located near to the disk center, $\approx$N00E18. The flare events took place at the southern part of the active region indicated by the circle.  Figure 2 displays the profiles of GOES soft X-ray flux, extreme ultra-violet (EUV) flux from SOHO/SEM \cite{judge1998}, and H$\alpha$ relative intensity of the flare events on the above date. The soft X-ray intensity profiles in the 0.5-4 \AA \ and 1-8 \AA \ bands, show three prominent consecutive flares, C3.8, M3.2, and M3.9, respectively, having peaks at 05:30, 07:52, and 08:31 UT. According to {\it Solar Geophysical Data} (SGD), two intense flares have been reported as a single event of 2N intensity in H$\alpha$, starting at 07:16 UT and a maximum at 07:54 UT. However, as seen in soft X-rays, the H$\alpha$ measurements clearly showed the progress of these two individual flares. It may be noted that the soft X-ray and H$\alpha$ flux profiles show two stages of energy release during the M3.2 flare event. Another interesting thing to note is that the M3.2 flare is more intense in H$\alpha$ in comparison to soft X-rays and EUV, whereas the M3.9 flare is more intense in soft X-rays and EUV in comparison to H$\alpha$.

\subsection{H$\alpha$ Observations}
The above flares were observed in H$\alpha$ at ARIES, Nainital, India, using
the 15-cm f/15 coud\' e solar tower telescope, equipped with a Bernard Halle 
H$\alpha$ filter and a Barlow-lens system to enlarge the image by a factor of two. 
The images were recorded by a 16-bit, 575$\times$384 pixel CCD camera system having 
a pixel size of 22 $\mu$m$^2$, which provides a resolution of 1$^{\prime\prime}$ per pixel
and a typical cadence of 15--20 s per image. 

\begin{figure}
\centering
\includegraphics[width=0.45\textwidth]{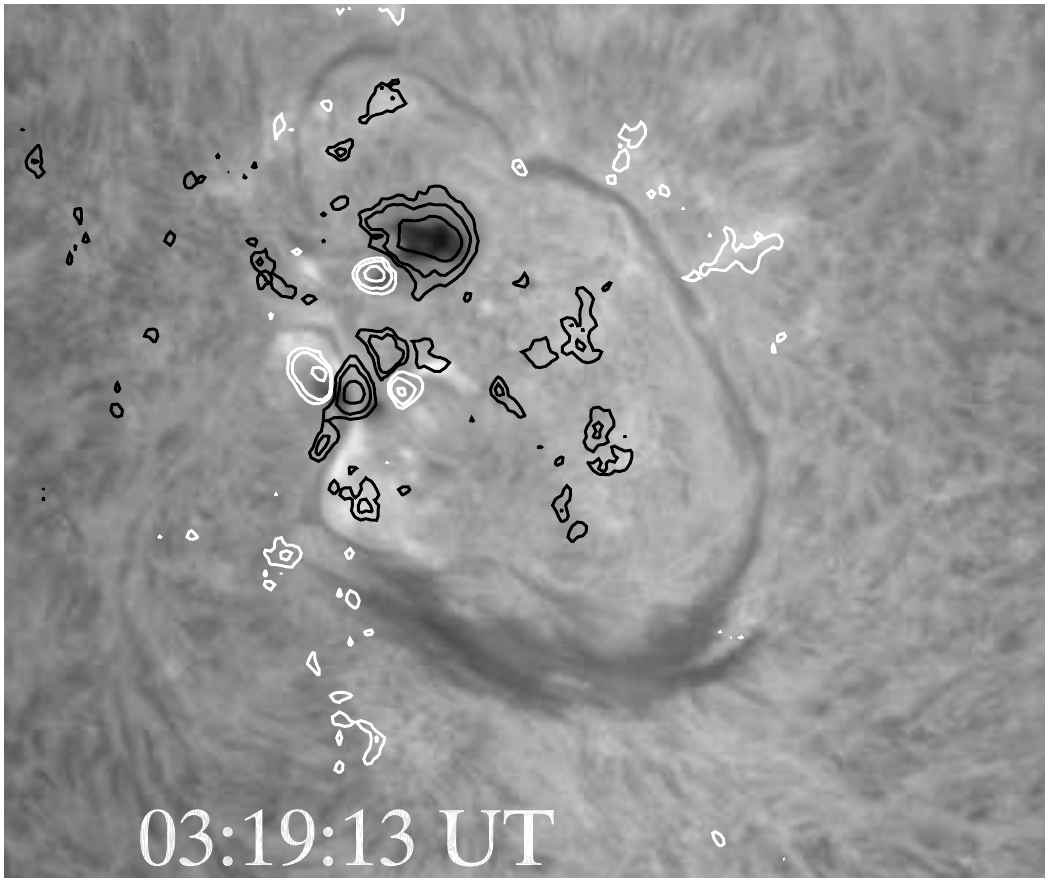}
\includegraphics[width=0.45\textwidth]{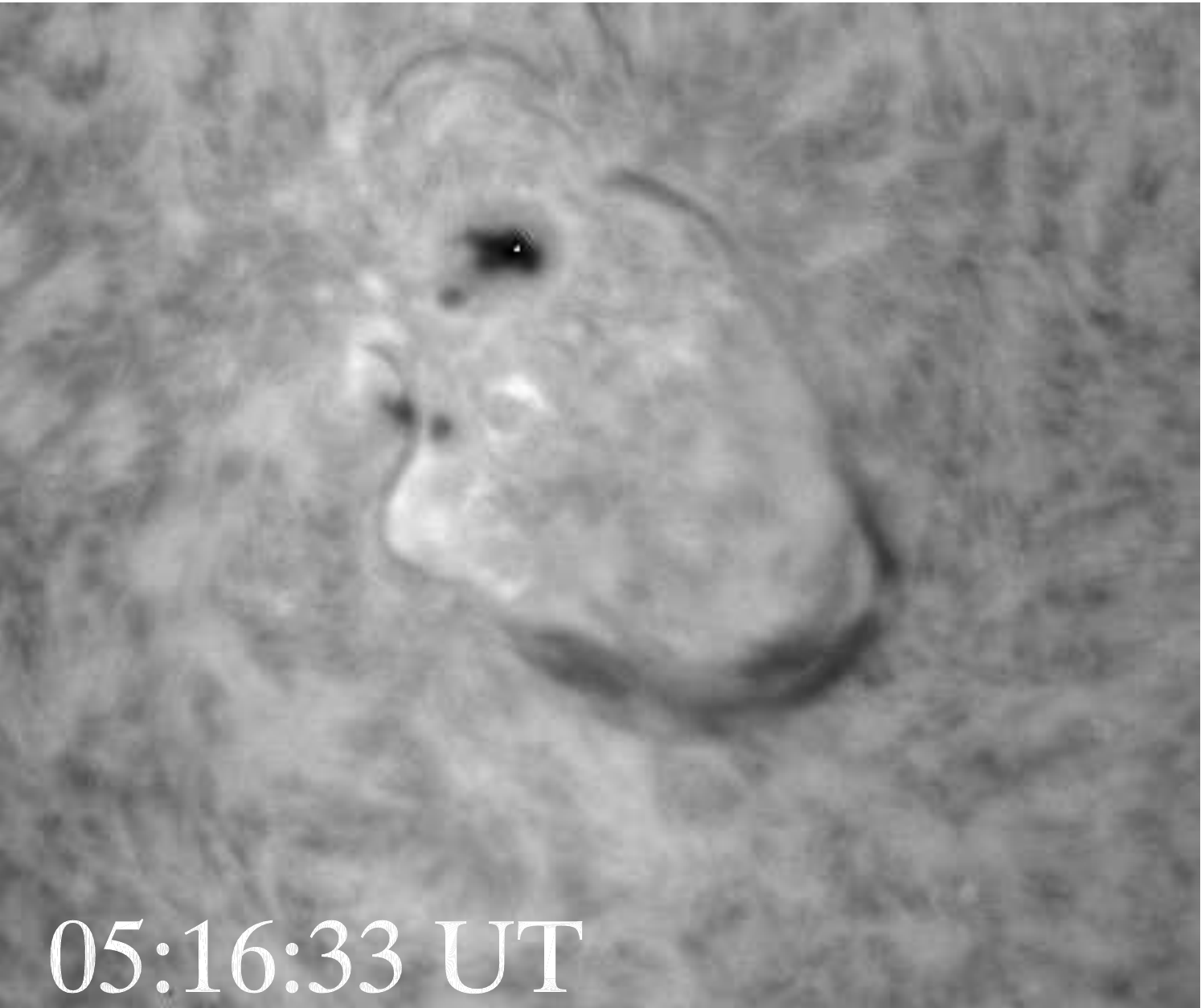}

\includegraphics[width=0.45\textwidth]{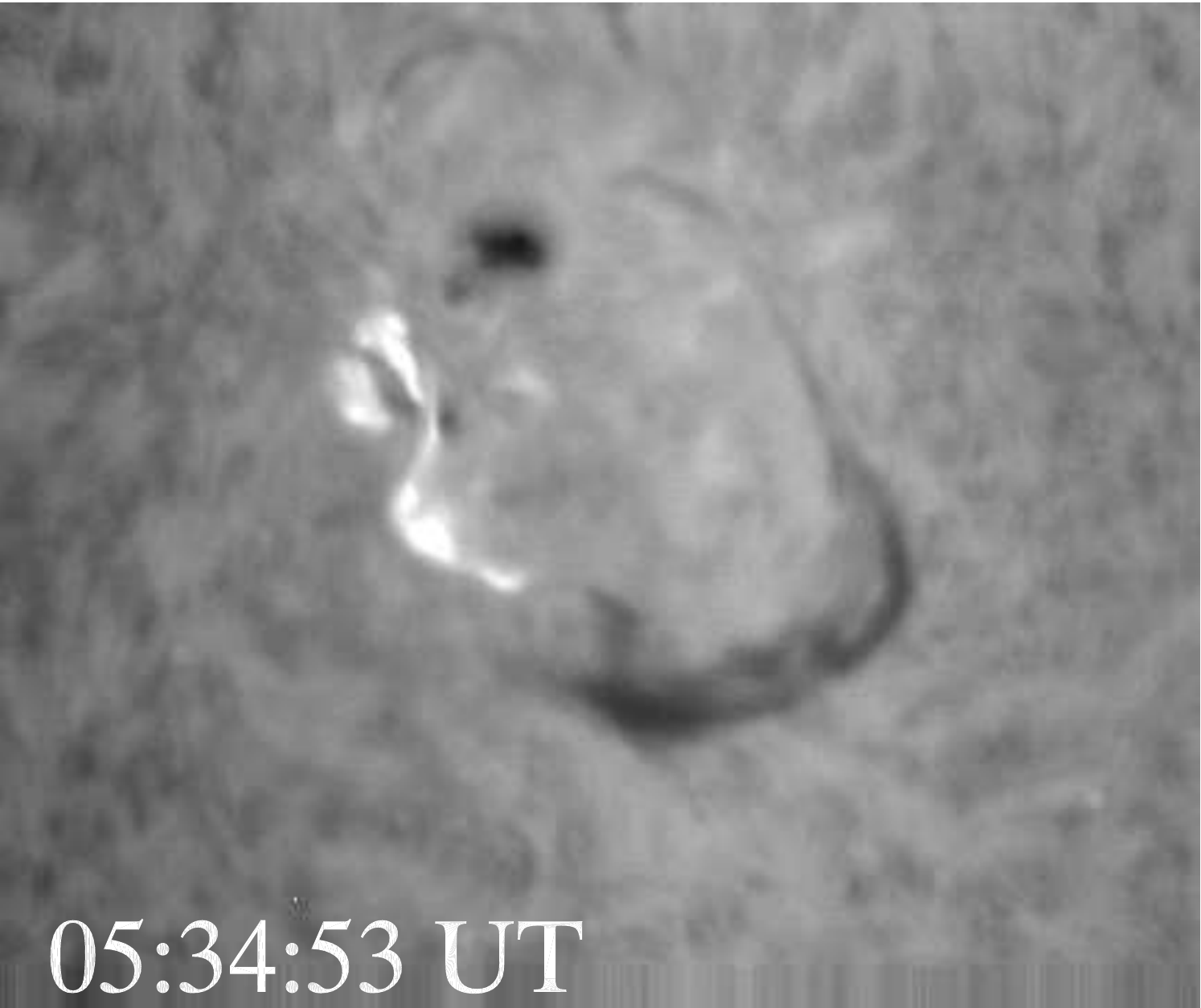}
\includegraphics[width=0.45\textwidth]{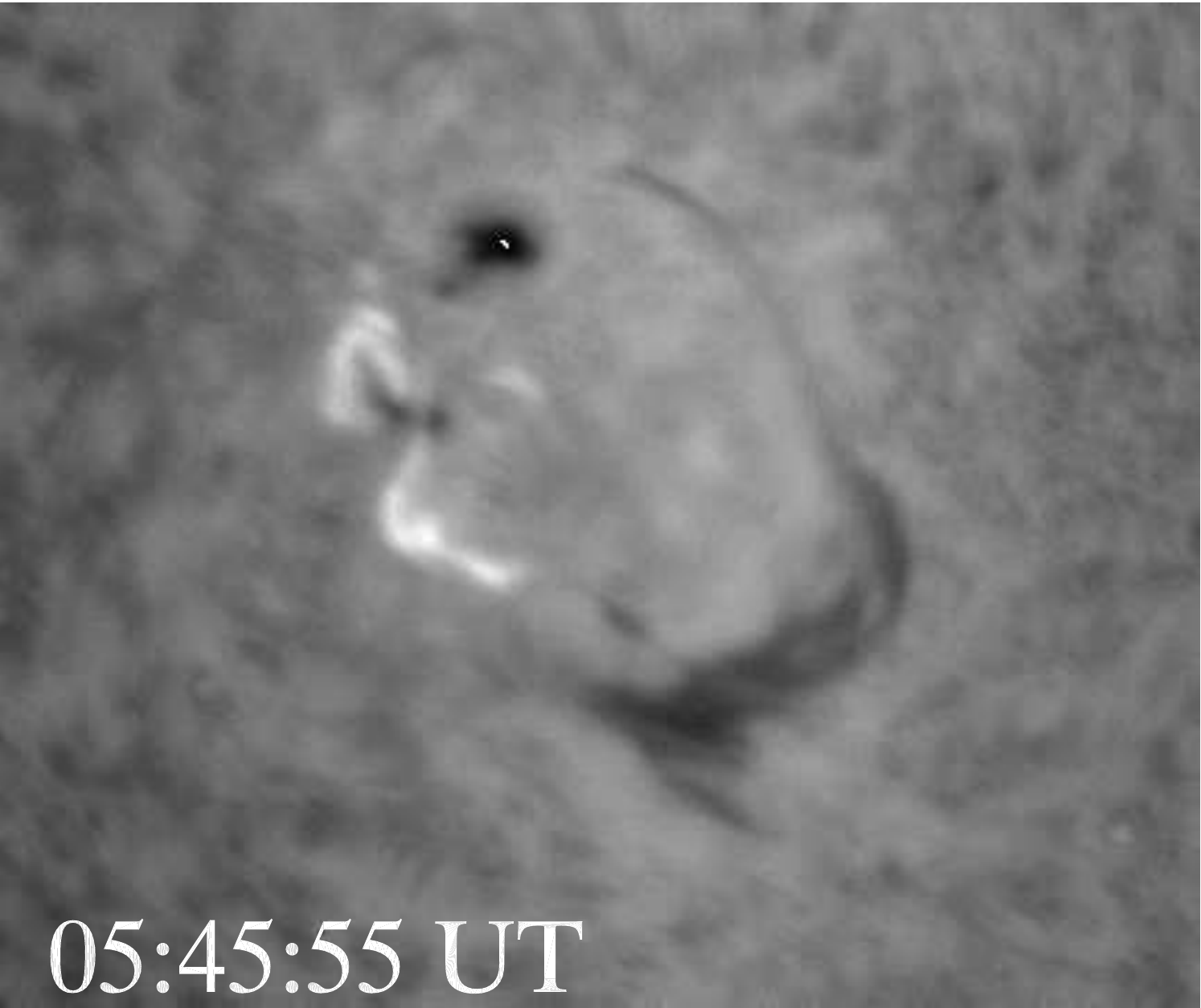}

\caption{H$\alpha$ images showing the evolution of the SF/C3.8 flare at the eastern footpoint of the filament. The top-left image is overlaid by SOHO/MDI magnetogram contours of positive (white) and negative (black) fields. The contour levels are $\pm$400, $\pm$800 and  $\pm$1600 G (gauss). The size of each image is 430$^{\prime\prime}$$\times$360$^{\prime\prime}$.}
\label{ha1}
\end{figure}
\begin{figure}
\centering

\includegraphics[width=0.45\textwidth]{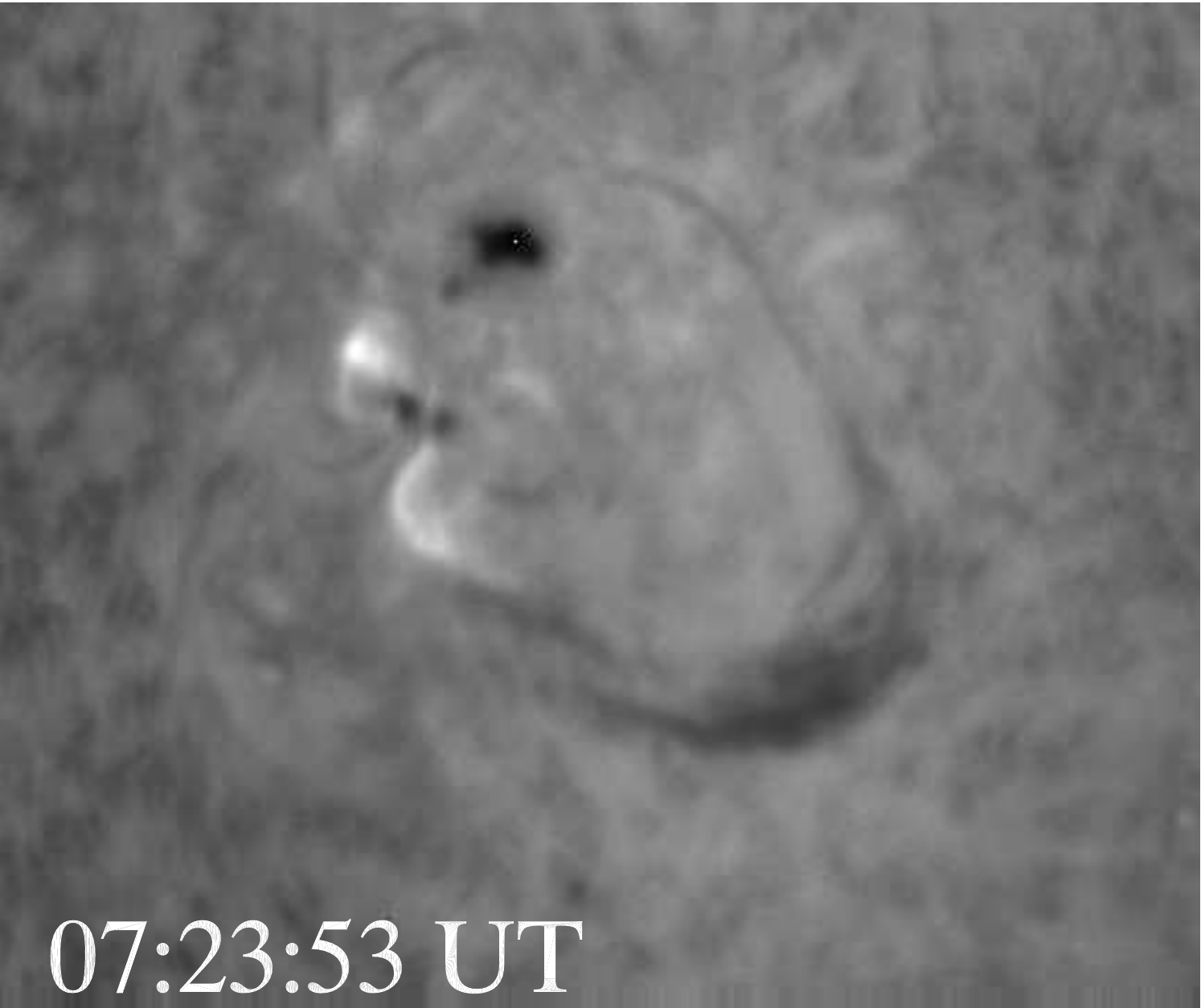}
\includegraphics[width=0.45\textwidth]{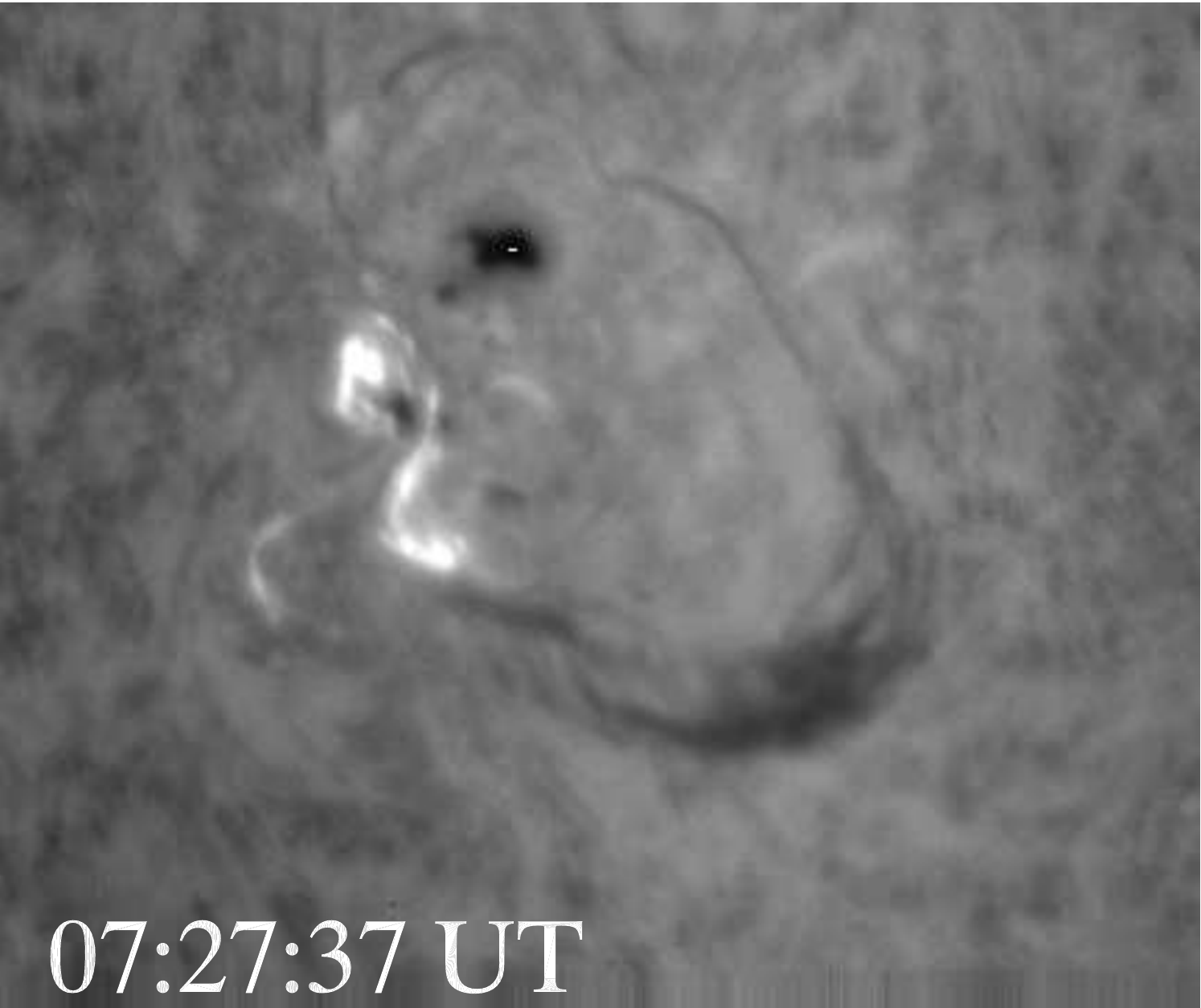}

\includegraphics[width=0.45\textwidth]{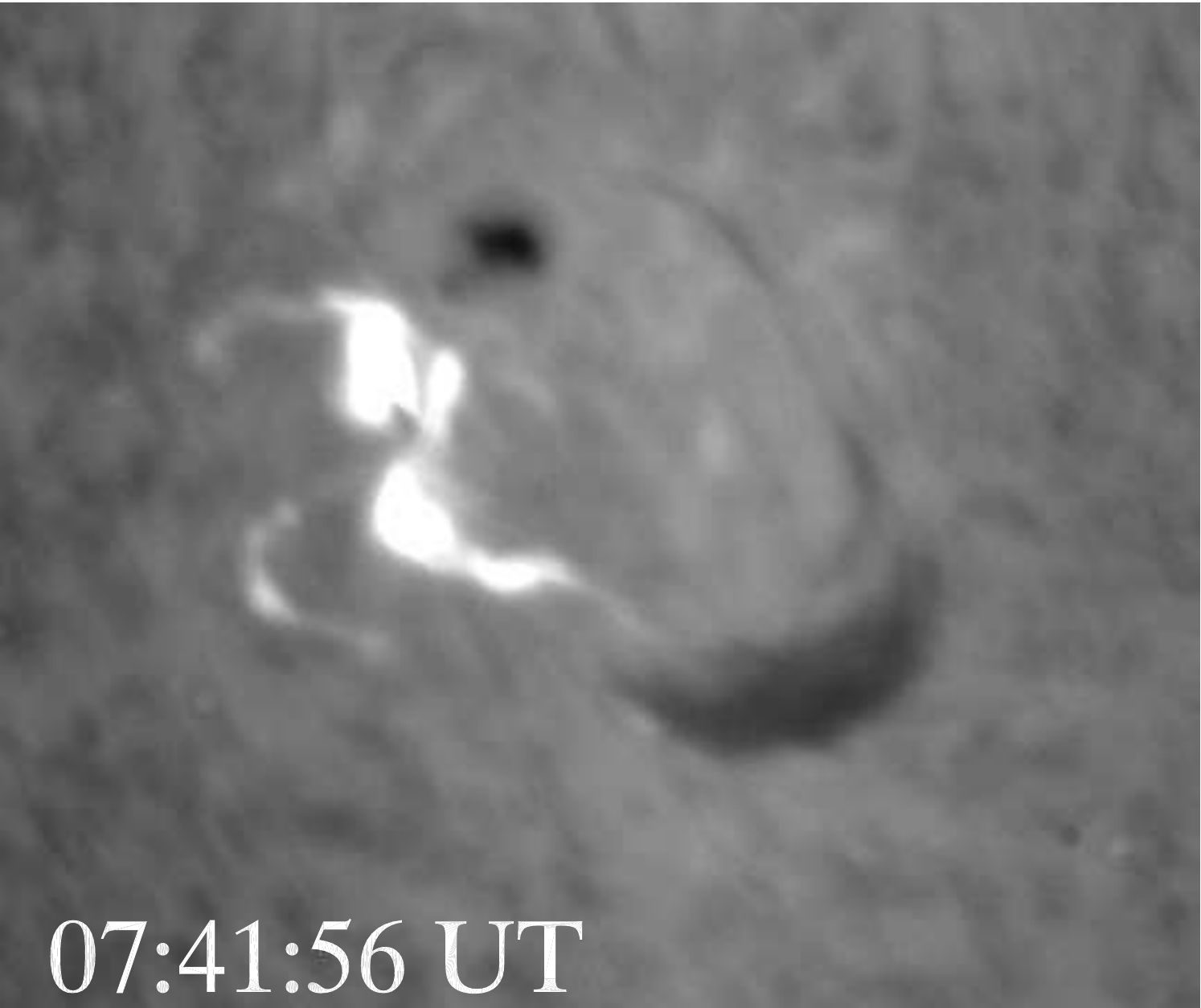}
\includegraphics[width=0.45\textwidth]{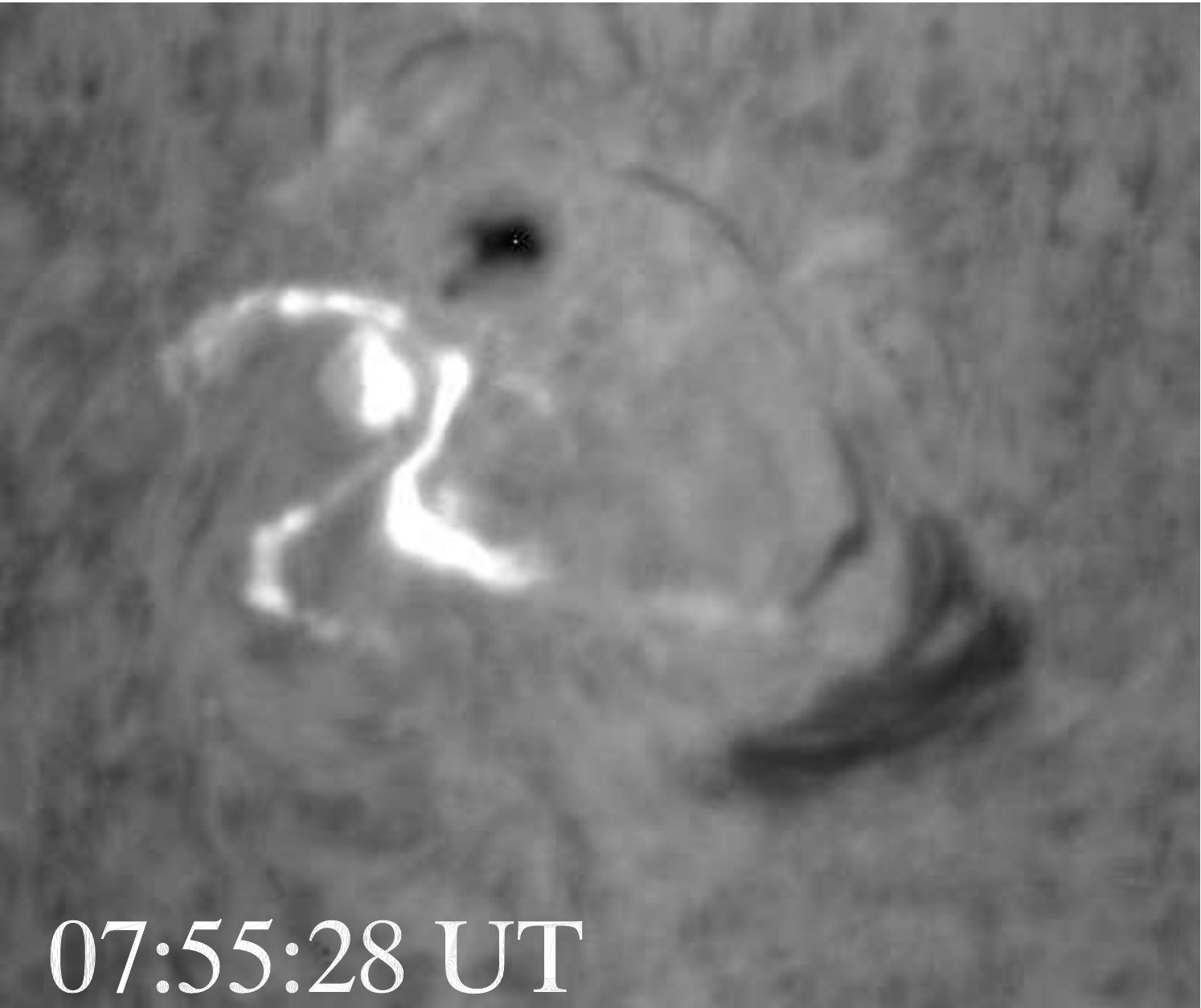}

\includegraphics[width=0.45\textwidth]{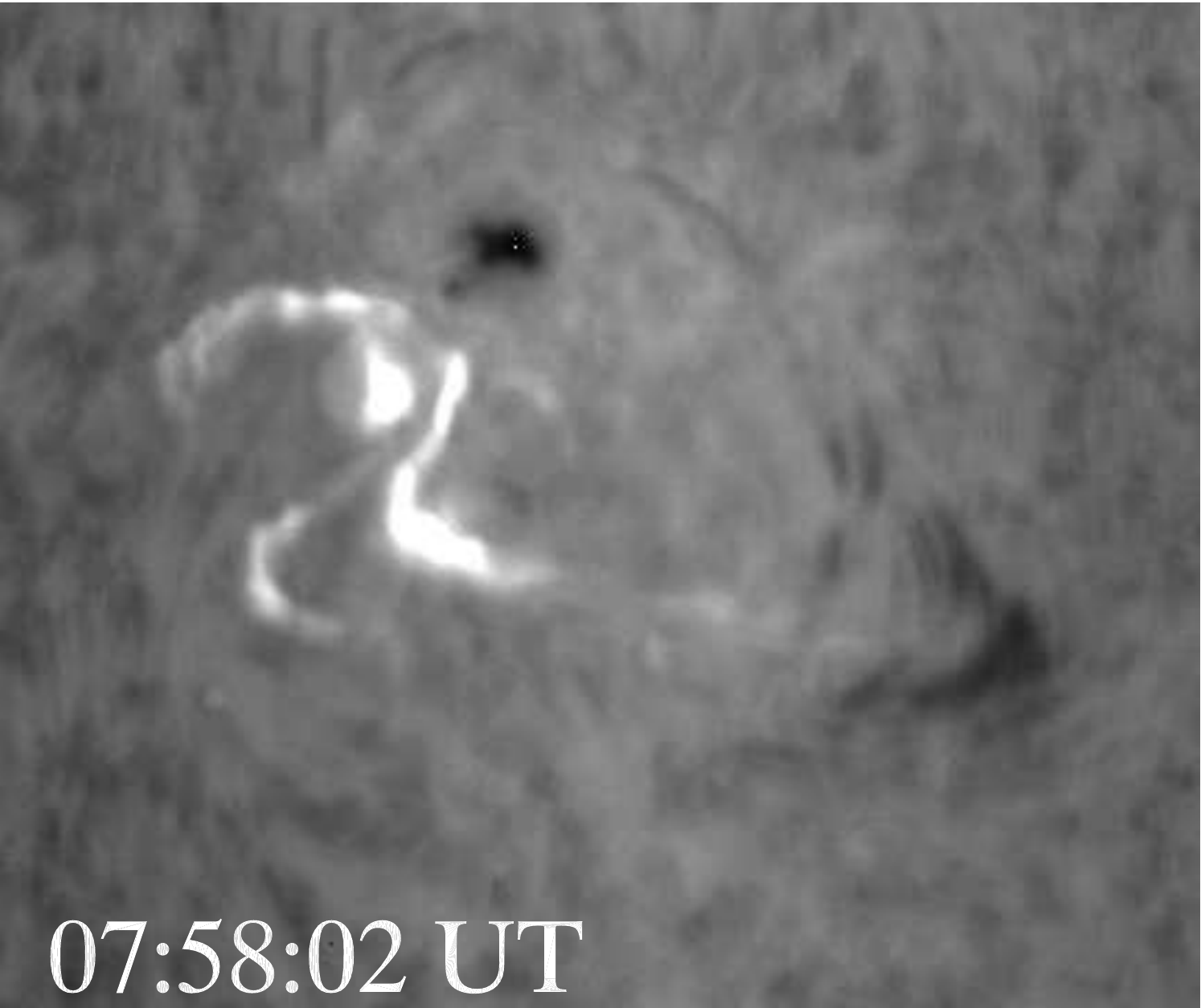}
\includegraphics[width=0.45\textwidth]{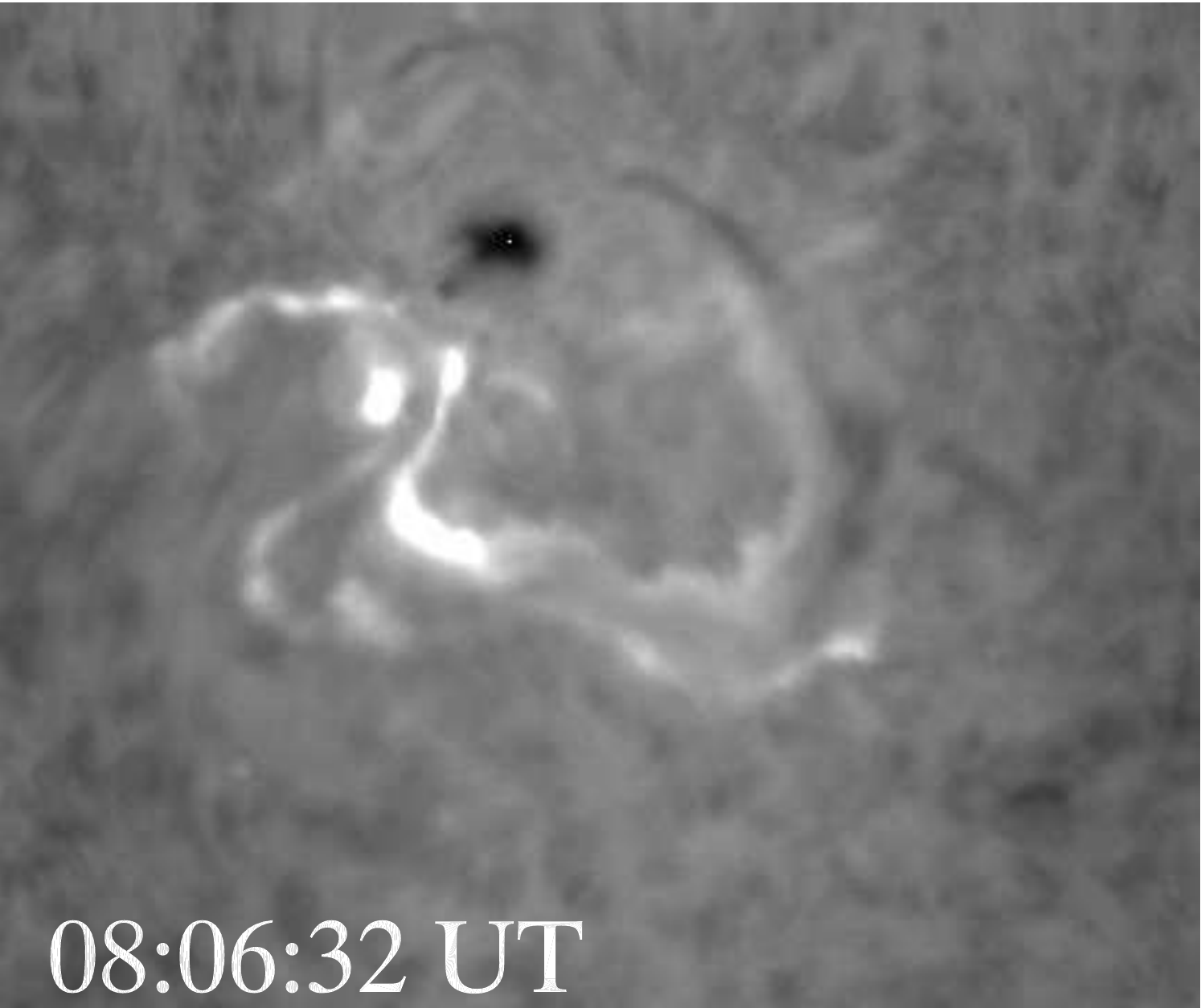}

\caption{H$\alpha$ images showing the evolution of the 2N/M3.2 flare. The size of each image is 430$^{\prime\prime}$$\times$360$^{\prime\prime}$.}
\label{ha2}
\end{figure}
\begin{figure}
\centering

\includegraphics[width=0.45\textwidth]{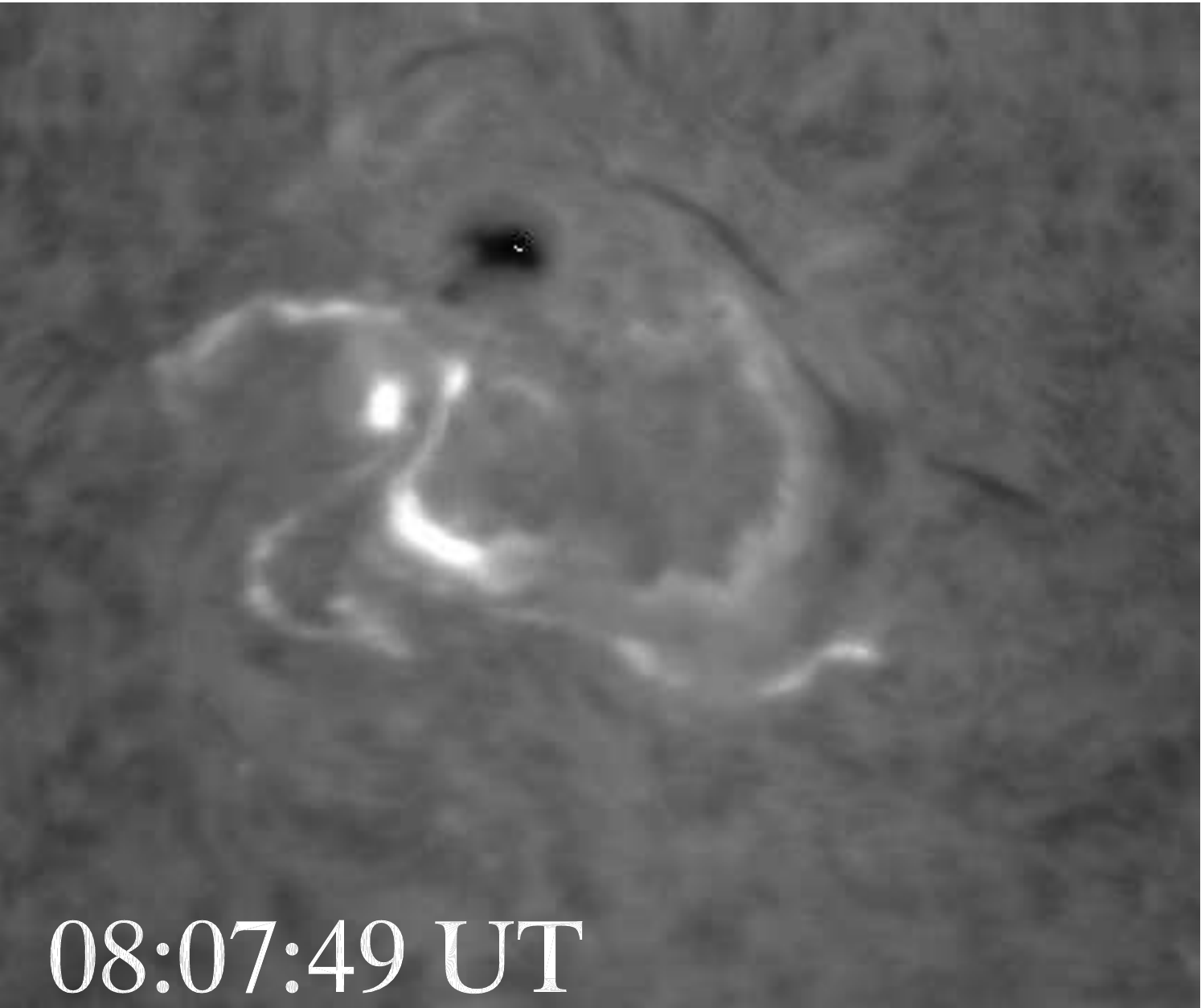}
\includegraphics[width=0.45\textwidth]{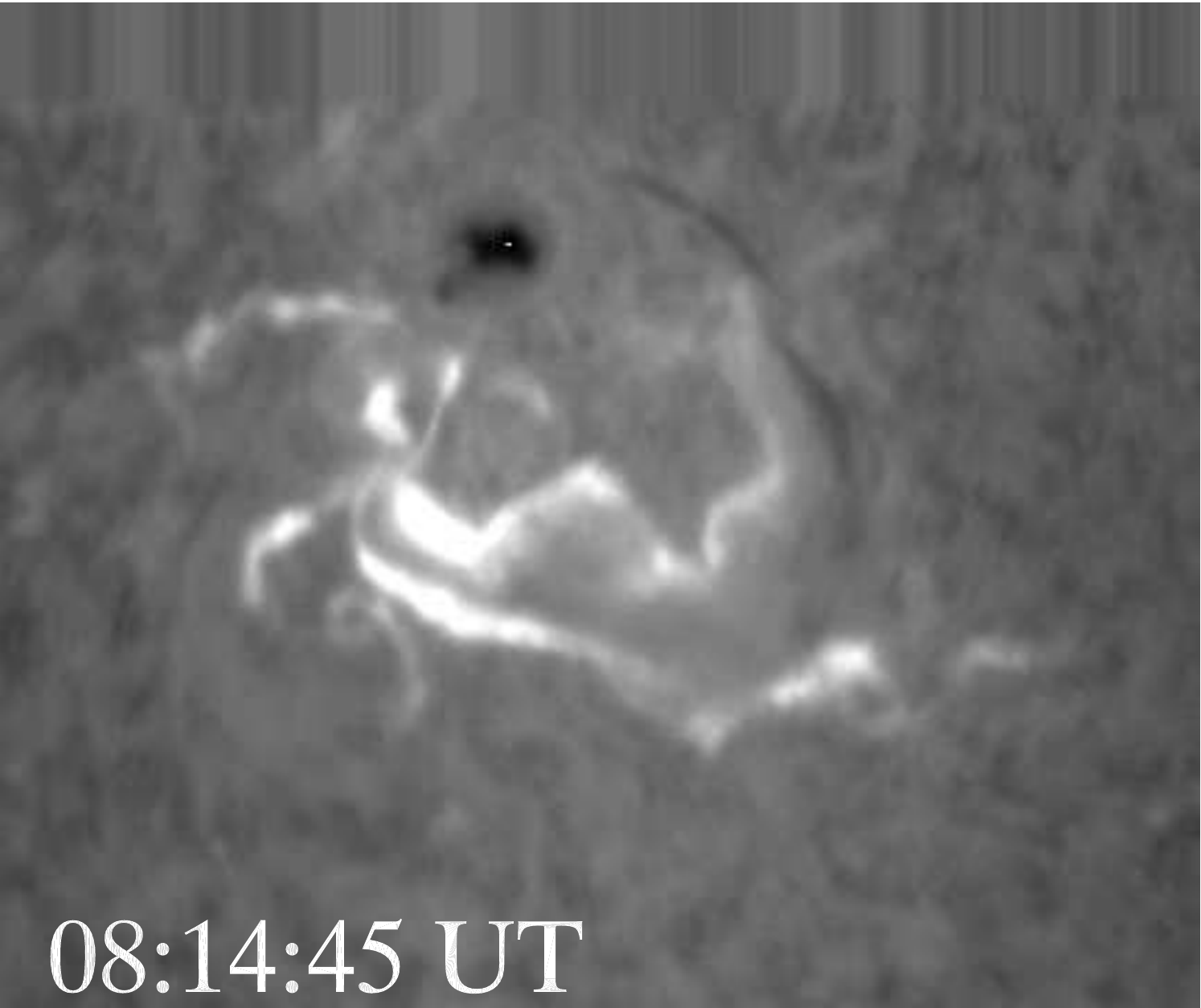}

\includegraphics[width=0.45\textwidth]{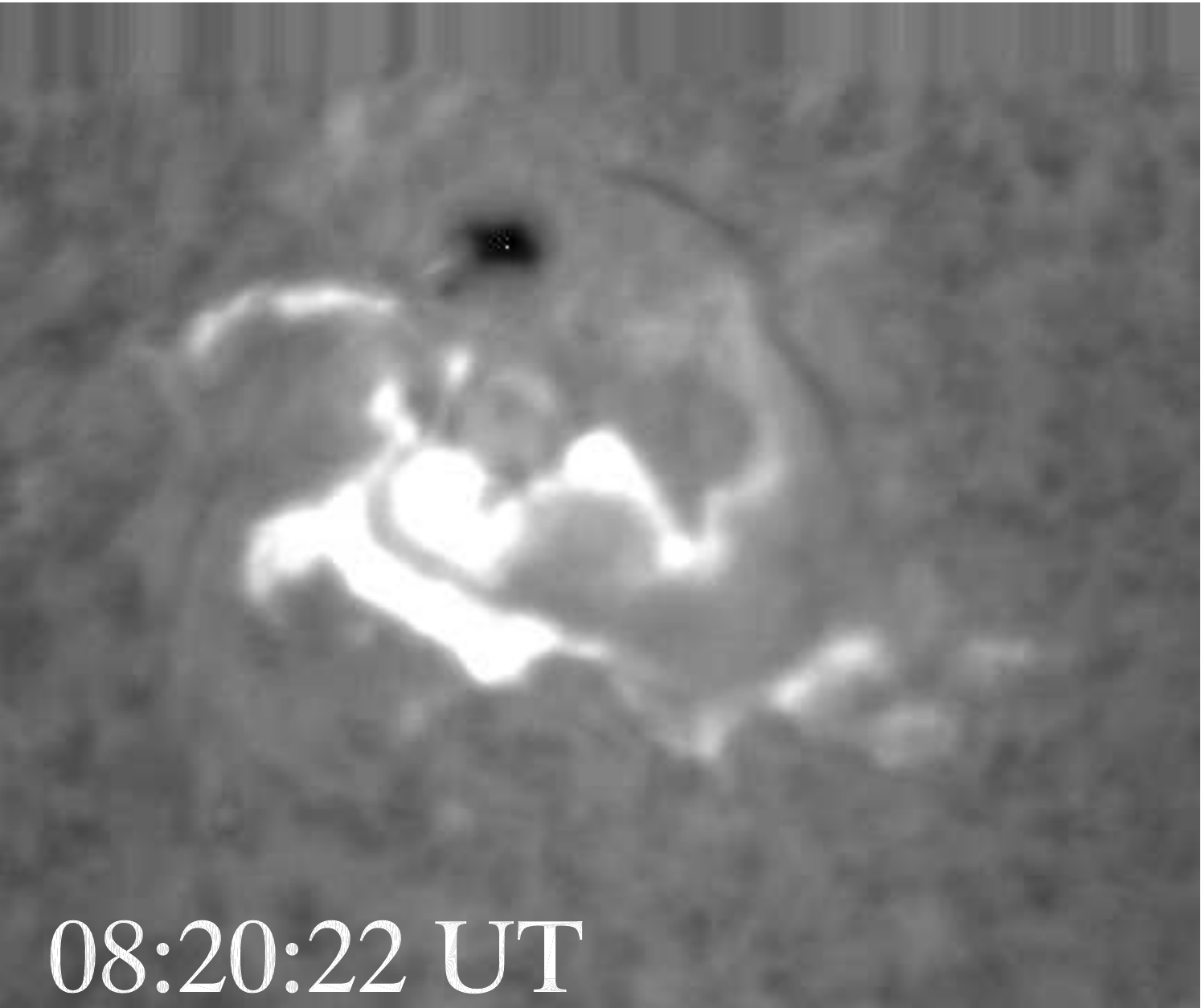}
\includegraphics[width=0.45\textwidth]{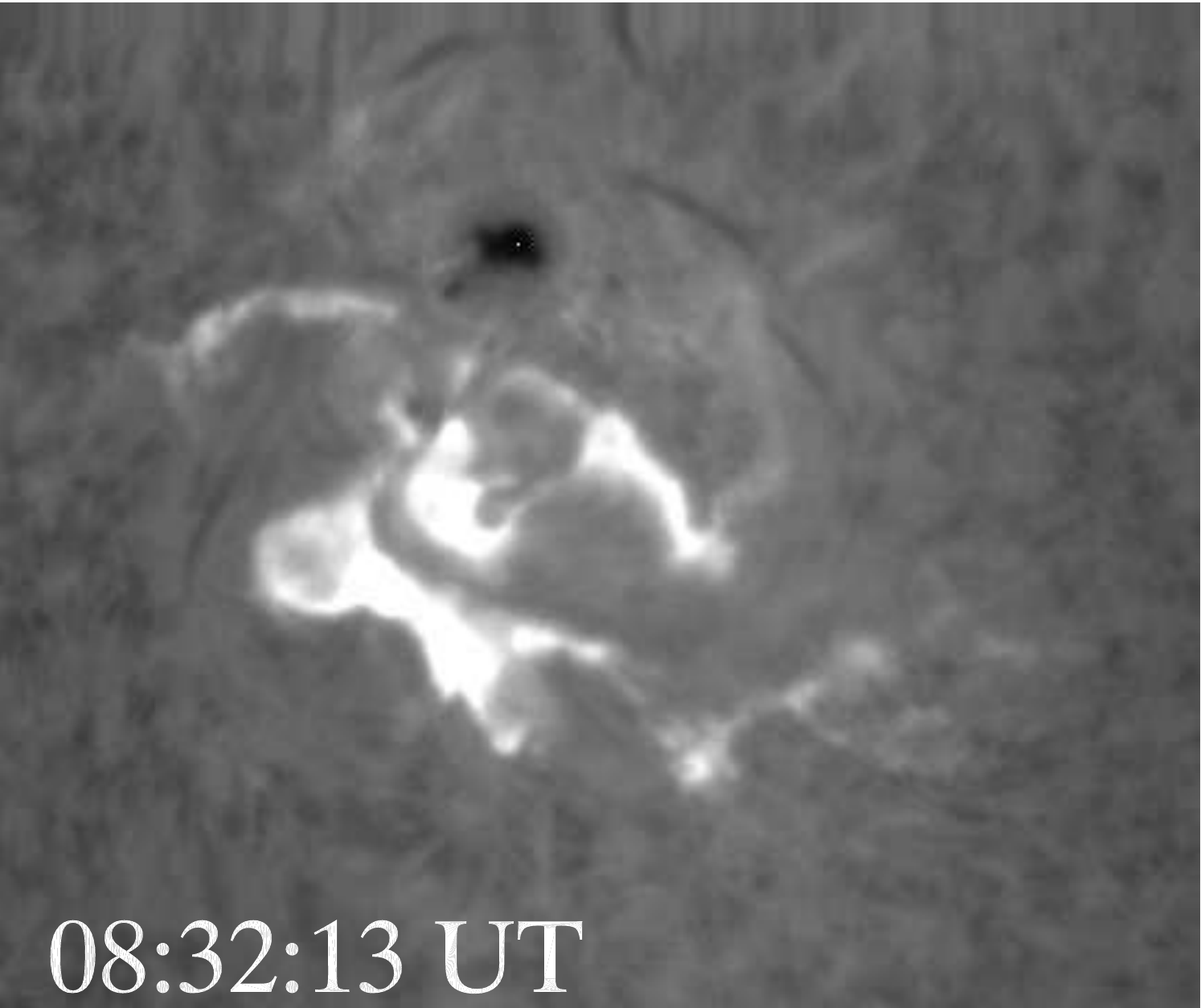}

\caption{H$\alpha$ images showing the evolution of the 2N/M3.9 flare. The size of each image is 430$^{\prime\prime}$$\times$360$^{\prime\prime}$.}
\label{ha3}
\end{figure}

Some of the selected H$\alpha$ images are shown in Figures \ref{ha1}, \ref{ha2} and \ref{ha3}, which reveal the evolution of C3.8, M3.2 and M3.9 flares respectively. The images taken before the flare onset at 03:19 UT show a huge U or horse-shoe shaped filament along with plage 
brightening near the sunspots and reveal the complex structure of the active region. Further, two channels of the filament can be seen at its south-eastern part. The C3.8 flare started with the heating and brightening at the south-eastern footpoint of the filament, and during the flare event, the filament seemed to get detached from this footpoint at 05:30 UT.  The curved  filament at the eastern foot point got activated during this flare event and a part of it erupted before the peak of the second flare M3.2. The remaining filament material moved away from the active region, between 07:48 and 08:10 UT, with a speed of $\approx$100 km s$^{-1}$. The curved filament almost disappeared before the onset of the third flare, M3.9. The flare onset at the location of high magnetic shear was shown by two flare ribbons with heavy wringles. As in the case of typical two-ribbon flares, these ribbons moved away from each other. The ribbon separation speed in H$\alpha$ was $\approx$44 km s$^{-1}$ . During the flare, the H$\alpha$ ribbons showed twisted structure, indicating a high level of shear. When the flare was in the decay phase, the flare ribbons returned to a simple parallel structure which revealed the relaxation of shear from non-potential to potential magnetic fields.

The eastern leg of the filament showed a highly sheared configuration before the onset of the second flare M3.2, and suggested a high level of twist at the location of the filament.  As the  flare progressed, the filament was destabilized and detached from the flare site.
The H$\alpha$ profile clearly showed that the relative intensity of the peak of M3.2 event was larger than that of the M3.9 flare. It suggests that the reconnection and heating associated with the M3.2 flare was taking place in the lower atmosphere where most of the energy was released as well as particles were accelerated during the filament eruption (refer to Figure \ref{fluxes}). The reconnection site is most likely at the eastern leg of the filament where two oppositely directed field lines reconnected, leading to ejection of mass, followed by the tether-cutting process. As the filament moved away from the active region, the H$\alpha$ brightness was enhanced rapidly. This filament system was possibly destabilized or reconnected with the overlying magnetic field in the corona, as it moved upward leading to the initiation of the intense X-ray flare of M3.9. It is likely that this M3.9 flare took place at a higher altitude than the previous one, as suggested by H$\alpha$ data. The top left panel of Figure \ref{ha1} displays the MDI contours overlaid  on the H$\alpha$ images prior to the flare event (left panel) to see the basic morphology of the magnetic field. In this plot, black contours show negative polarity whereas white ones indicate positive polarity. 

\subsection{TRACE EUV and GOES/SXI Images}
\begin{figure}
\hspace{8.5mm}
\includegraphics[width=0.3\textwidth]{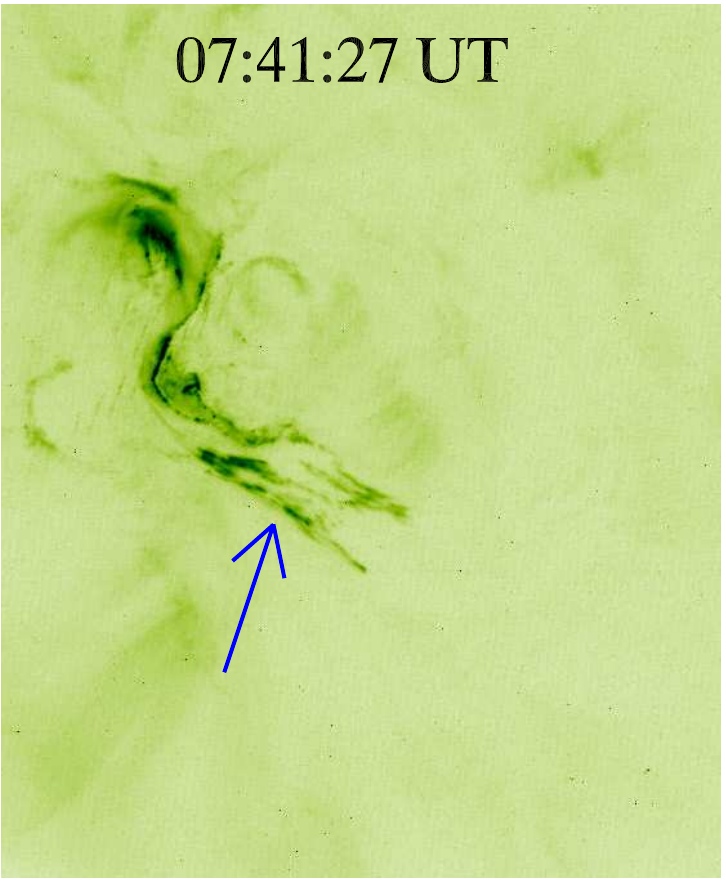}
\includegraphics[width=0.3\textwidth]{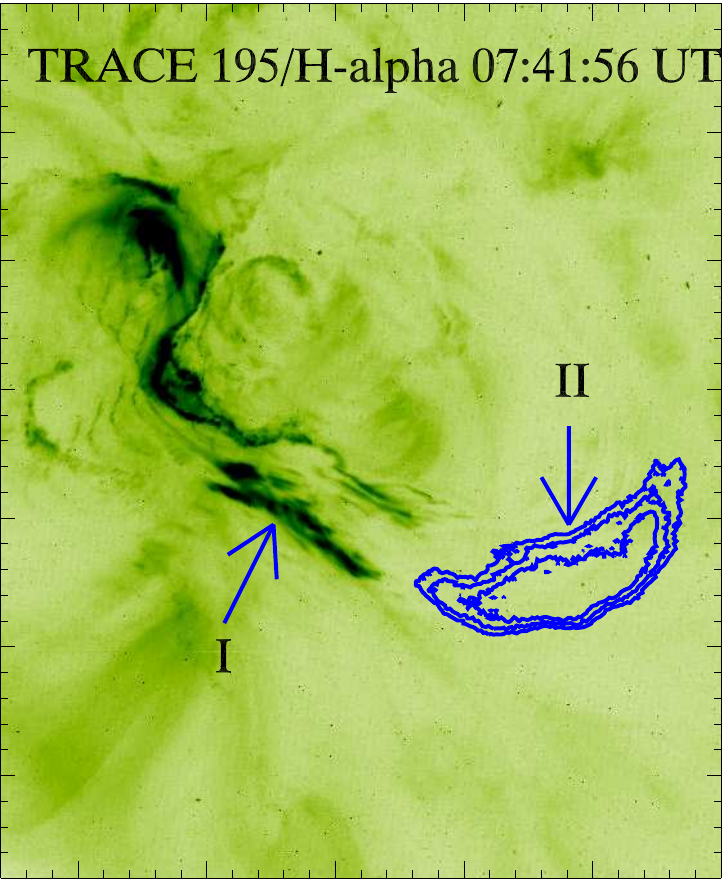}
\includegraphics[width=0.3\textwidth]{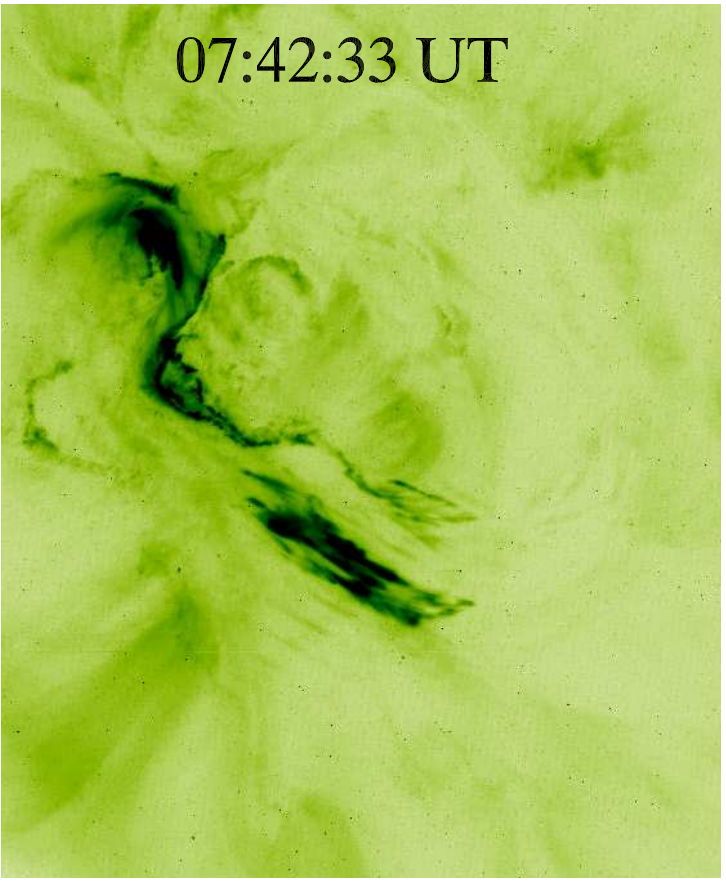}

\hspace{8.5mm}
\includegraphics[width=0.3\textwidth]{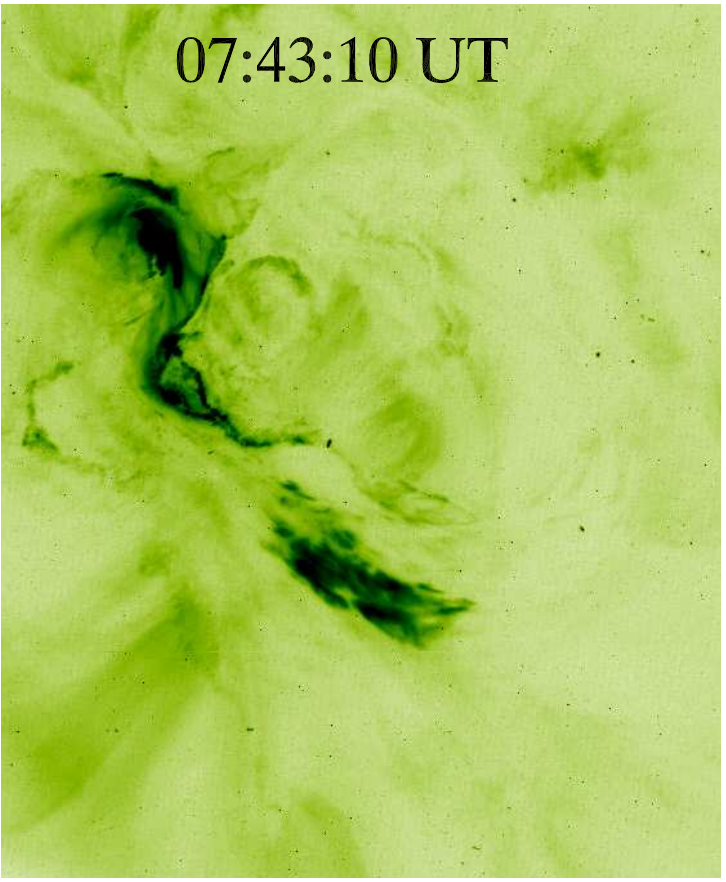}
\includegraphics[width=0.3\textwidth]{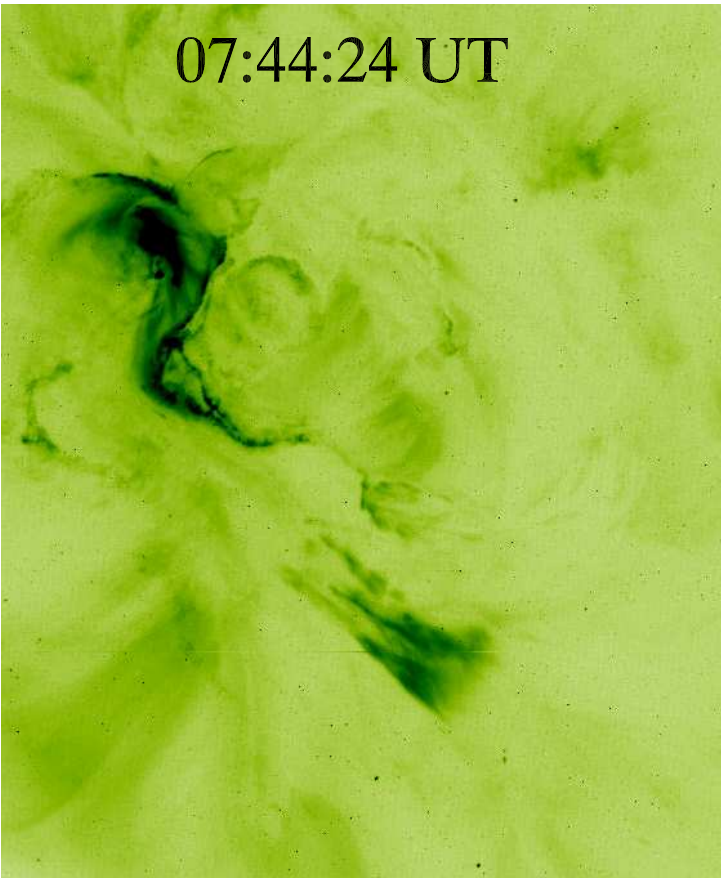}
\includegraphics[width=0.3\textwidth]{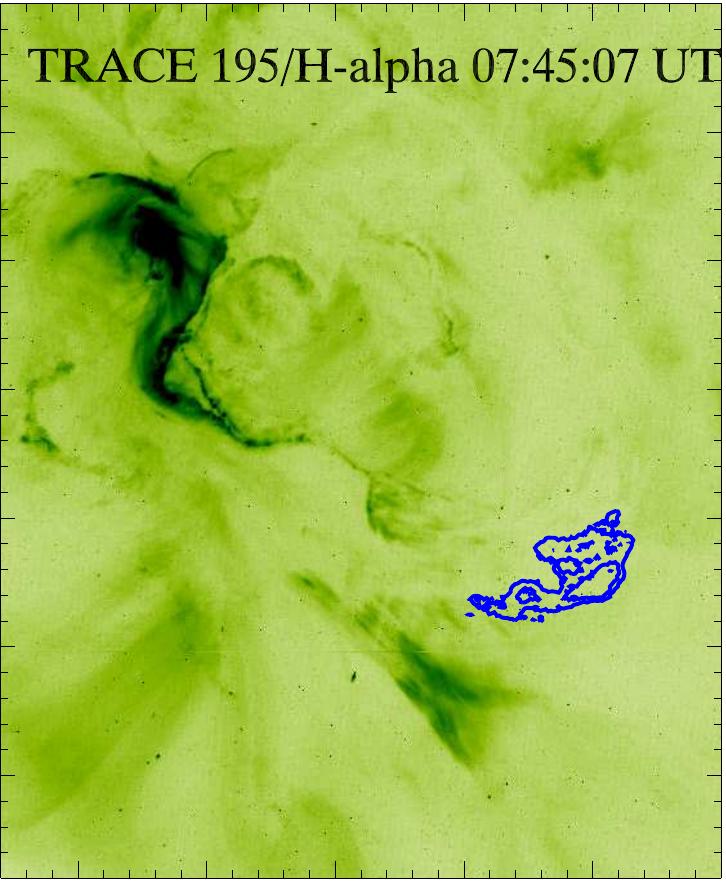}

\hspace{2.4mm}
\includegraphics[width=0.35\textwidth]{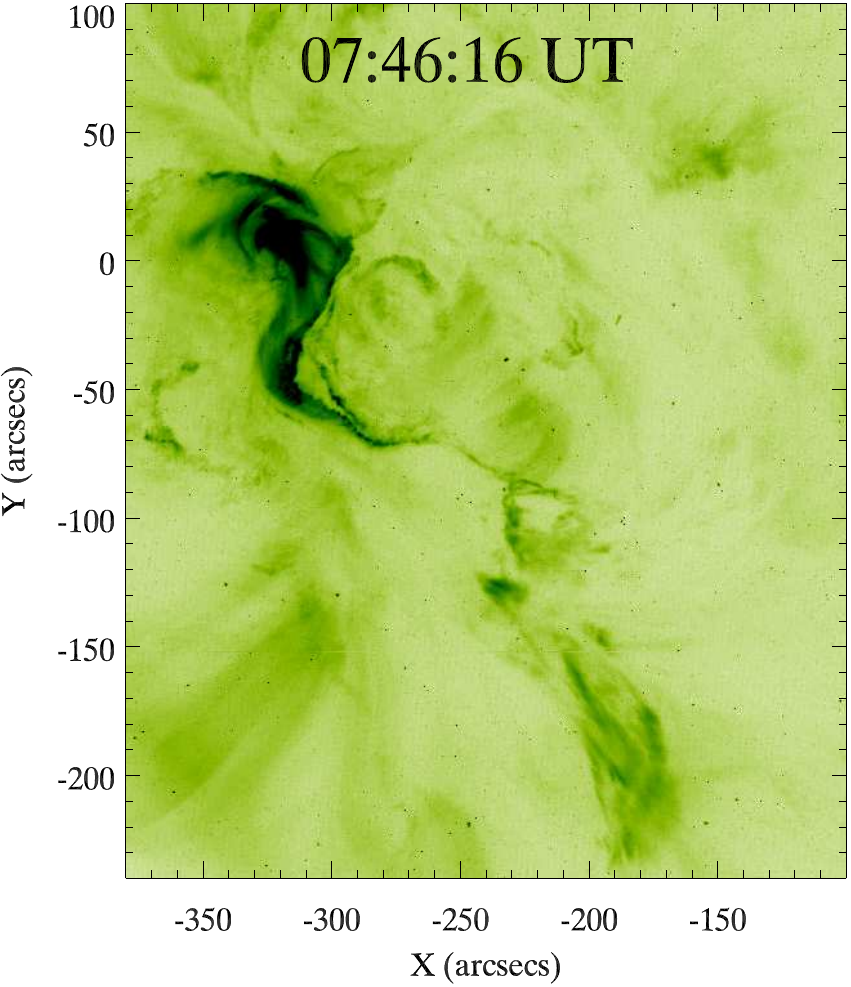}
\includegraphics[width=0.3\textwidth]{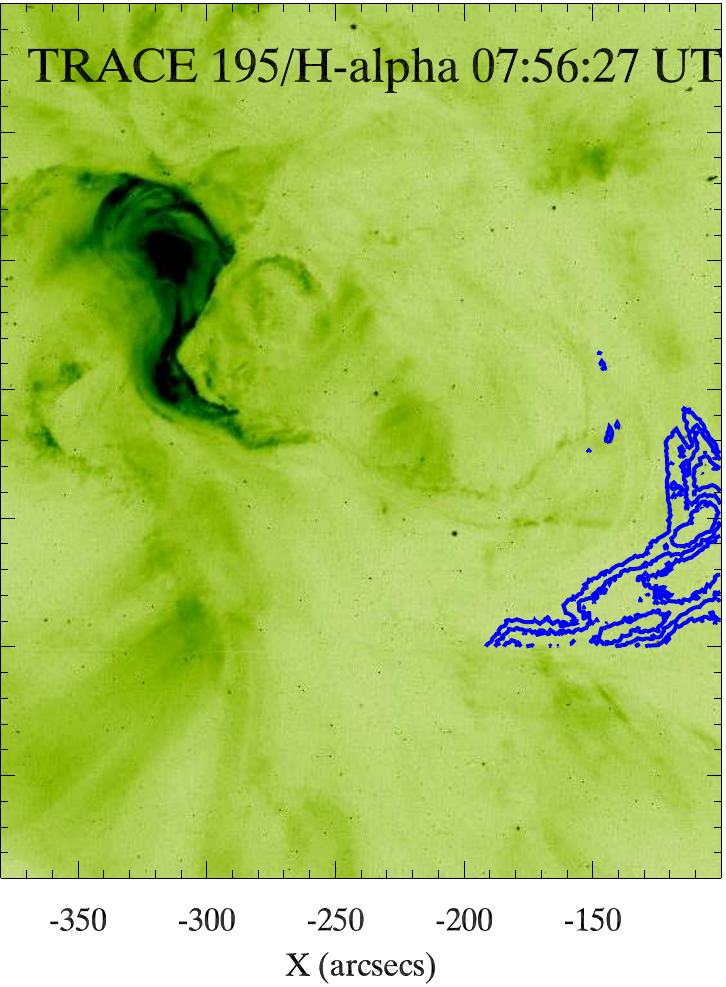}
\includegraphics[width=0.3\textwidth]{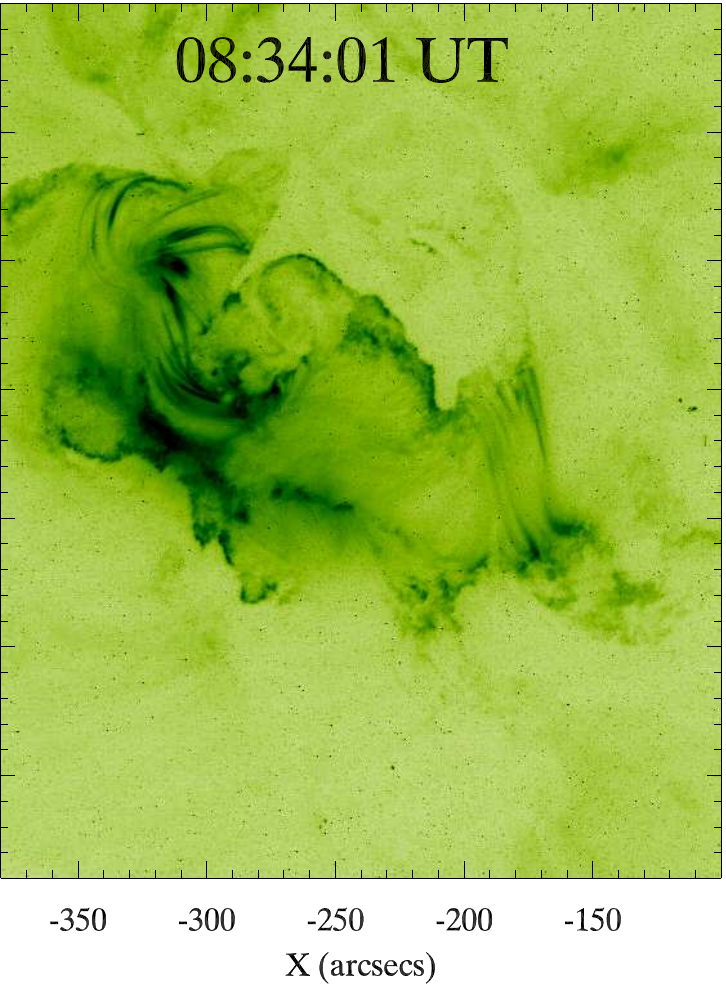}

\caption{ TRACE 195 \AA \ (Fe {\sc xii} line) negative images showing the flare evolution with the filament eruption (top row, left). The image at 07:41:56 (top row, middle) shows the H$\alpha$ filament contours overlaid on  the TRACE 195 \AA \ image. This image confirms the eruption of two different channels of the same filament (first visible in EUV and second one in H$\alpha$, indicated by arrows). The image at 07:56:27 UT (bottom row, middle) shows the presence of the second filament channel in H$\alpha$ while the first filament channel moved away from TRACE field of view.}
\label{tr195}
\end{figure}

The evolution of the filament eruption with successive flare events has also been studied with the help of TRACE 195 \AA \ images \cite{handy1999}, which correspond to $\approx$1.5 MK plasma (Fe {\sc xii} line). Figure \ref{tr195} displays the sequence of selected TRACE images
which show the flare and the filament eruption during 07:41--07:47 UT. We have overlaid the H$\alpha$ filament contours over the selected TRACE images of the same time (07:41, 07:45 and 07:56 UT). At 07:41 UT, the filament channel (indicated by I) started to erupt at its eastern footpoint and moved away from the AR (shown by arrow). It was observed up to 07:48 UT during the M3.2 flare event. The estimated speed of the filament was  $\approx$330 km s$^{-1}$. The flare intensity enhanced as the filament moved up and the flare maximum took place at 07:52 UT and decayed after $\approx$08:06 UT.  

\begin{figure}
\centering
\includegraphics[width=0.45\textwidth]{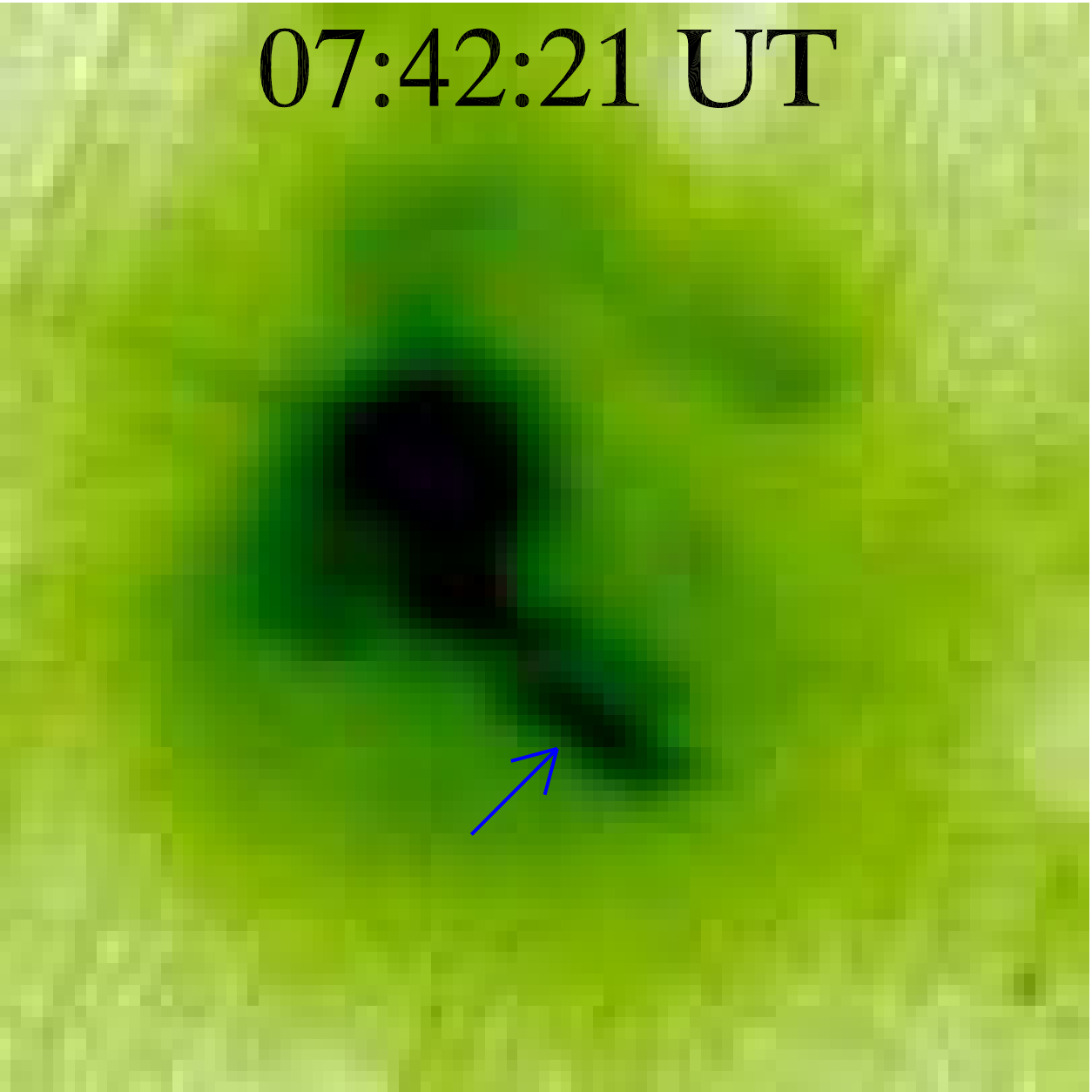}
\includegraphics[width=0.45\textwidth]{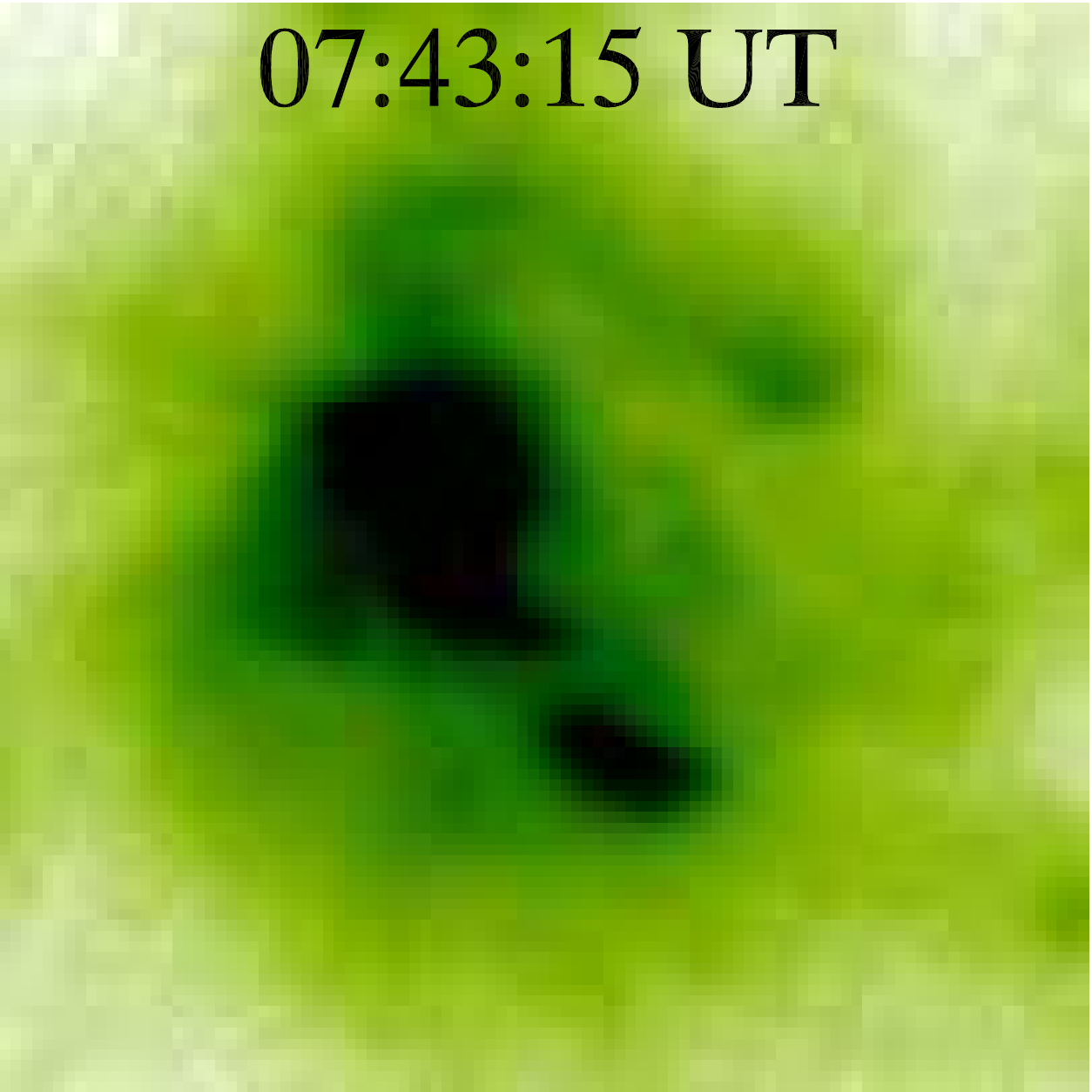}

\includegraphics[width=0.45\textwidth]{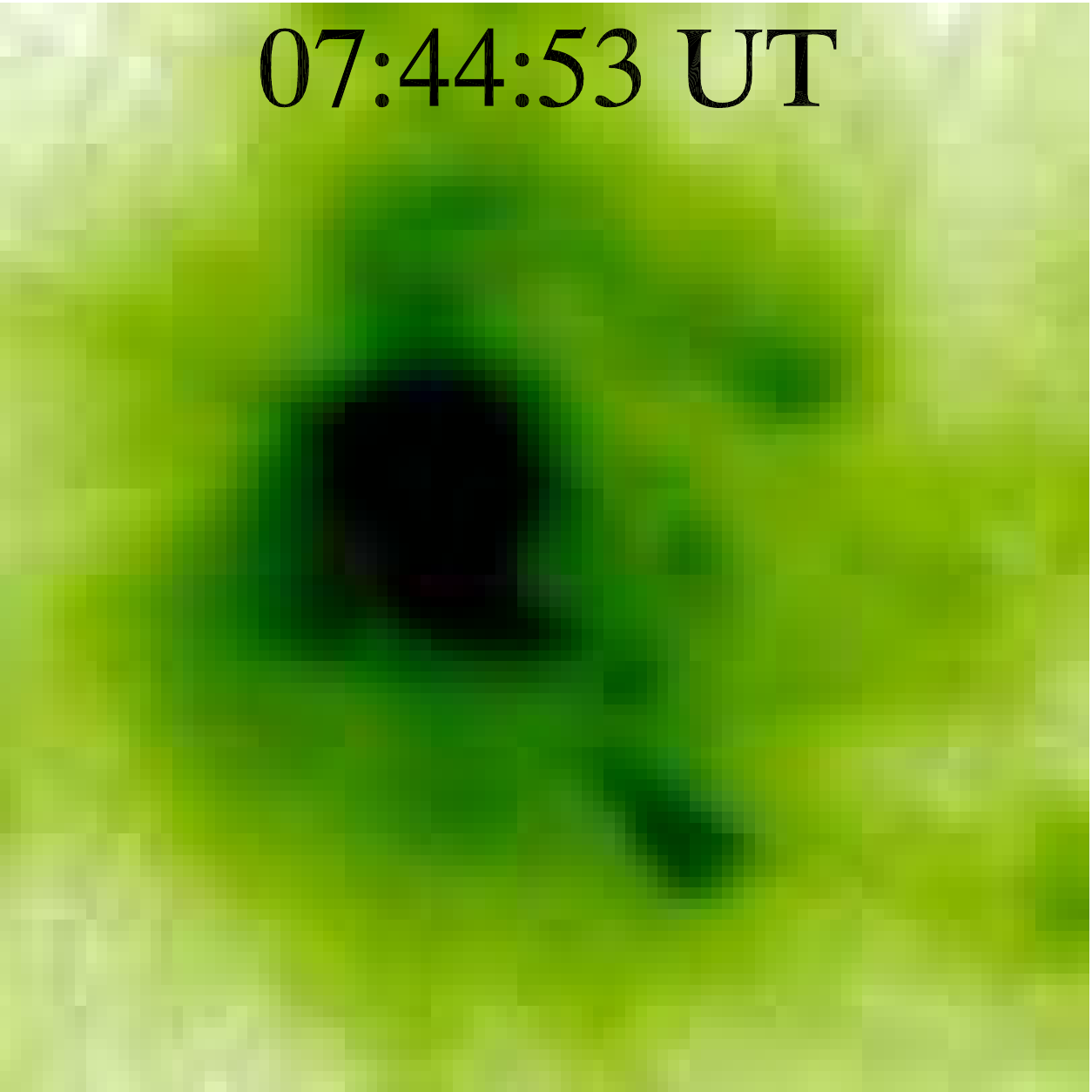}
\includegraphics[width=0.45\textwidth]{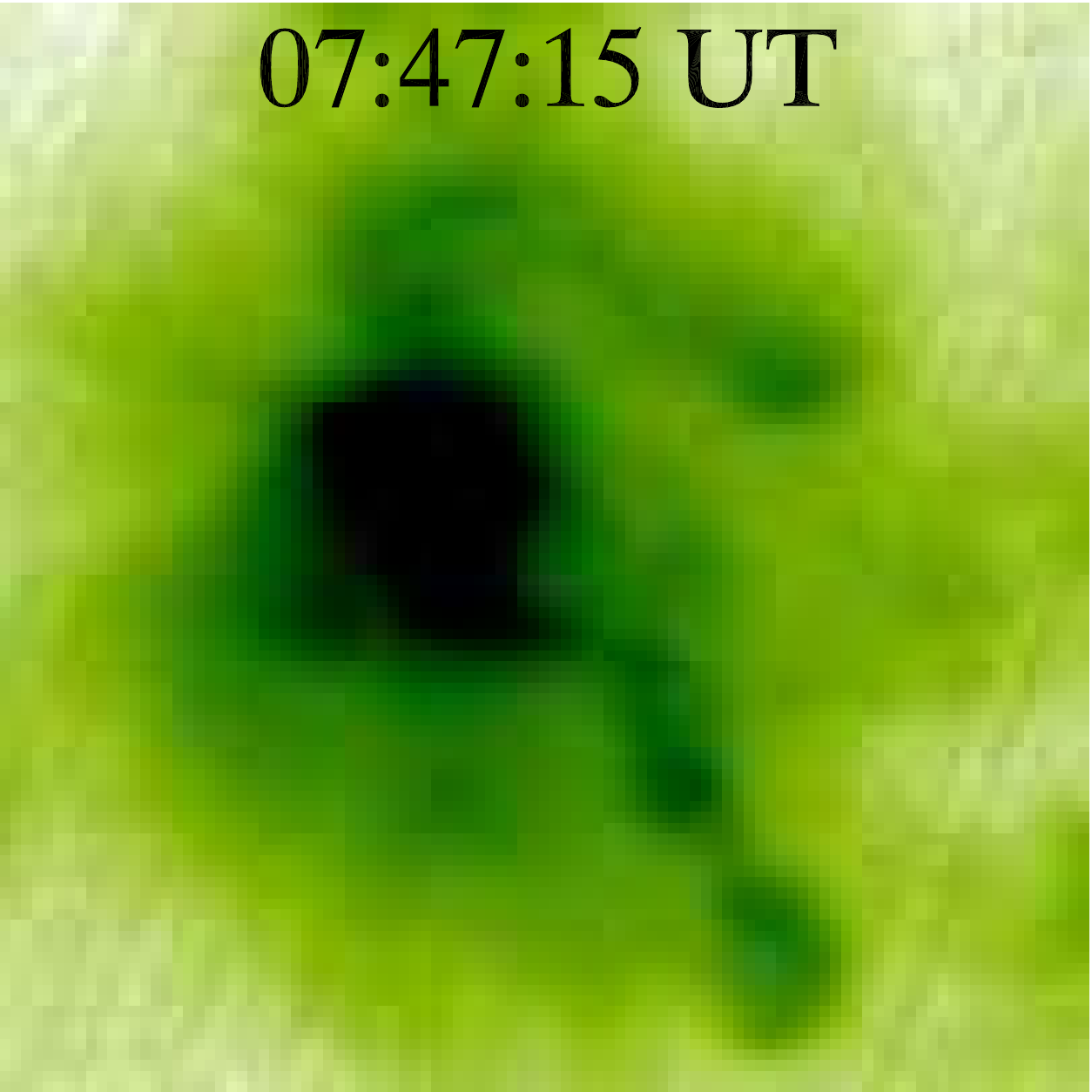}

\caption{GOES soft X-ray (SXI) images showing the filament eruption (indicated by arrow) associated with the second M3.2 flare. The size of each image is 500$^{\prime\prime}$$\times$500$^{\prime\prime}$.}
\label{sxi}
\end{figure}
We have used GOES-SXI observations of the event \cite{hill2005,pizzo2005}.
It is a broadband imager in the 6-60$^\circ$ \AA \ bandpass that produces full-disk solar images with $\approx$1 min cadence. The images consist of 512$\times$512 pixels with 5$^{\prime\prime}$ resolution. It observed both the flares and filament eruptions (Figure \ref{sxi}). The filament was seen at 07:42 UT and moved away from the active region, which was visible up-to 07:48 UT. We measure the projected speed of the filament using a linear fit to the height-time data points. The linear speed of the filament ejection was $\approx$ 370 km s$^{-1}$, which is in good agreement with the TRACE measurements. The H$\alpha$ filament contour overlaid on the TRACE image at 07:41 clearly confirms the presence of two filament channels/parts (indicated by I and II). The image at 07:56 suggests the presence of the central part of the H$\alpha$ filament (II) whereas the first filament channel, {\it i.e.} the eastern part of the filament, has moved away from the TRACE/SXI field of view. We compared these TRACE images with H$\alpha$ images and after a careful investigation, we found that the eastern part of the erupted filament was not visible in the H$\alpha$ wavelength. This may be due to its high temperature, which emitted radiation only in shorter wavelengths (TRACE and SXI).

\begin{figure}
\centering
\includegraphics[width=0.9\textwidth]{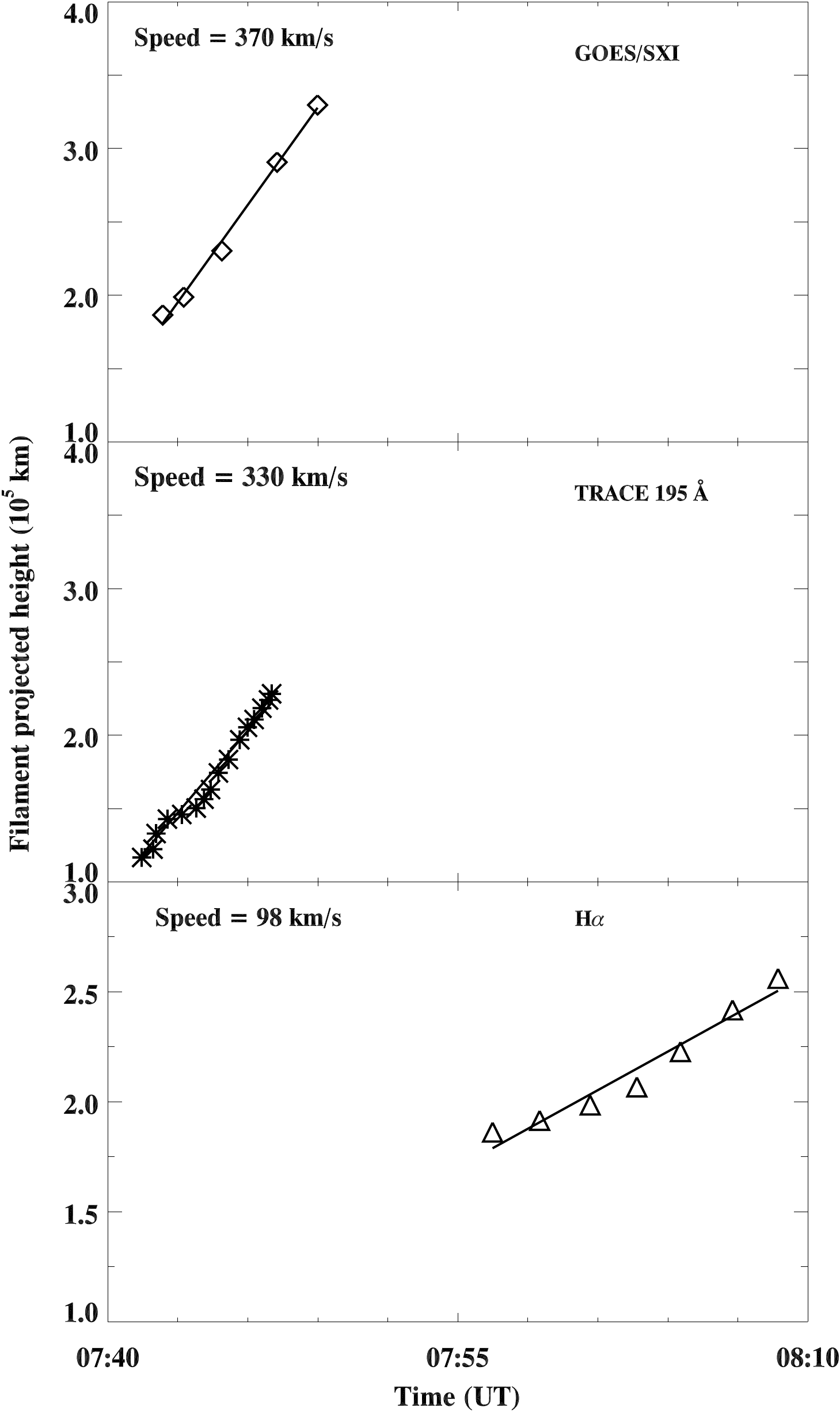}
\caption{The height-time profiles of the filament eruption in GOES SXI, TRACE 195 \AA \, and H$\alpha$ during the flare events.}
\label{fil_ht}
\end{figure}

 Figure \ref{fil_ht} displays the projected height-time profiles of the filament ejections recorded in TRACE, SXI, and H$\alpha$ observations. It is evident from the plot that there were two different ejections from the same filament moving with different speeds. From the linear fit to the data points, the first ejection had a very high speed ($\approx$330-370 km s$^{-1}$) whereas the second one had a speed of $\approx$100 km s$^{-1}$.

\begin{figure}
\centering
\includegraphics[width=0.9\textwidth]{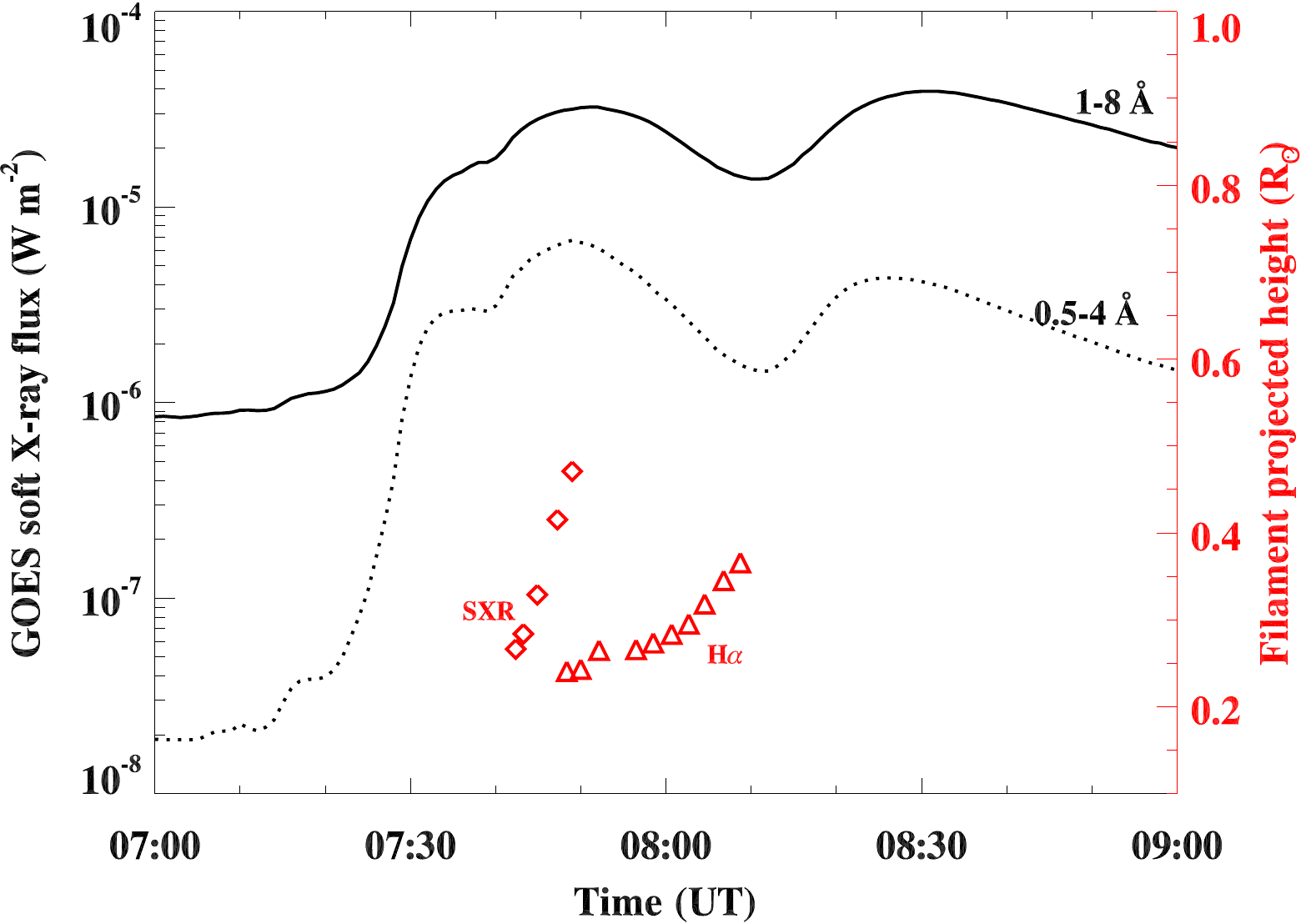}
\caption{GOES soft X-ray flux with the height-time profiles of the filament estimated from GOES SXR and H$\alpha$ images.}
\label{fil_xray}
\end{figure}
In Figure \ref{fil_xray}, two peaks are seen in soft X-rays during the M3.2 event, and these peaks indicate the energy release at two stages. It is likely that the energy release at the first stage of the M3.2 flare triggered the first ejection (eastern part). When this mass moved away from the active region, the flare intensity maximized around 07:40--07:50 UT and the second peak was attained, which is presumably due to the progressive reconnection with the overlying field lines caused by the ejection of the filament. As this CME moved out of the inner corona, another ejection got pushed from the chromosphere, in association with the M3.9 flare.

\begin{figure}
\centering
\includegraphics[width=0.9\textwidth]{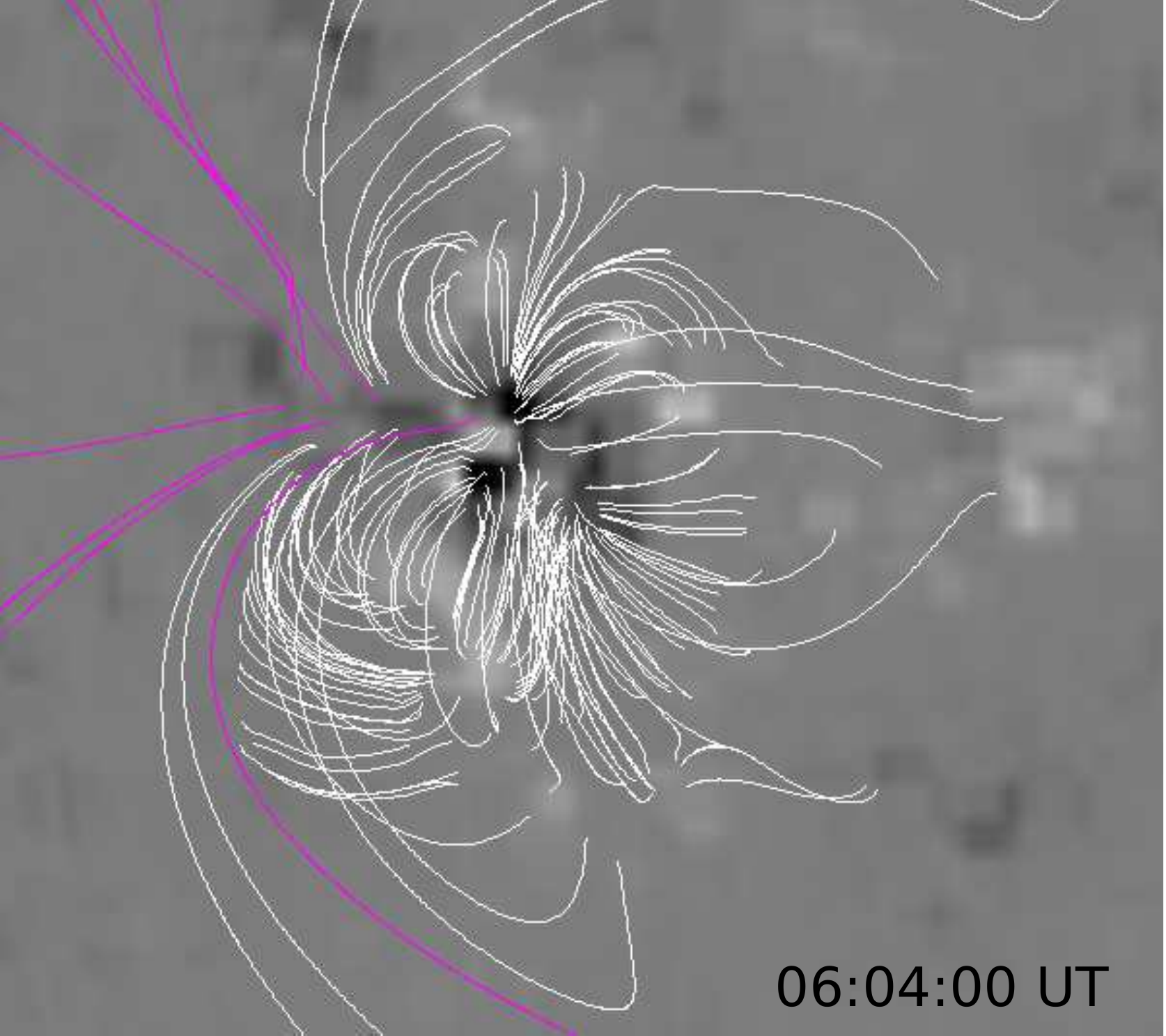}
\caption{Magnetic field lines by the PFSS exrapolation of the active region NOAA 10501 at 06:04:00 UT on 18 November 2003. The field of view of the MDI image is 800$^{\prime\prime}$$\times$750$^{\prime\prime}$.}
\label{pfss}
\end{figure}

 We have used the potential-field source-surface (PFSS) extrapolation \cite{alt1969,sch1969} before these flare events to understand the magnetic structure of the flare site. Figure \ref{pfss} displays the extrapolated field lines over the SOHO/MDI image of the active region. White lines show the closed field lines whereas red ones display the open field lines in the active region. Comparing with H$\alpha$ images, we may notice that the filament was located along the polarity inversion line covered with the low-lying field lines in the active region. The higher magnetic field lines also connect above the low-lying field lines.
\subsection{White Light Observations}
\begin{figure}
\centering

\includegraphics[width=0.45\textwidth]{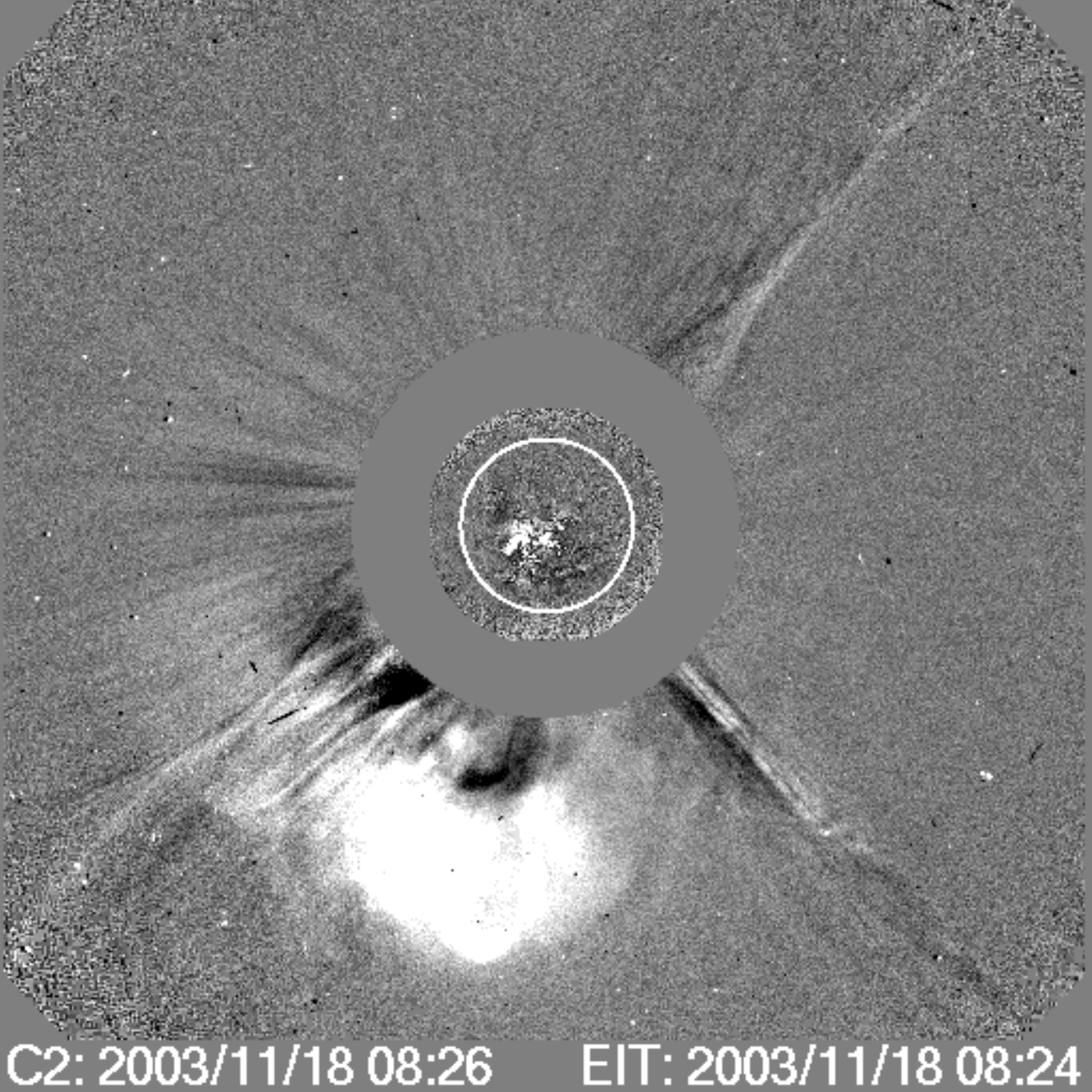}
\includegraphics[width=0.45\textwidth]{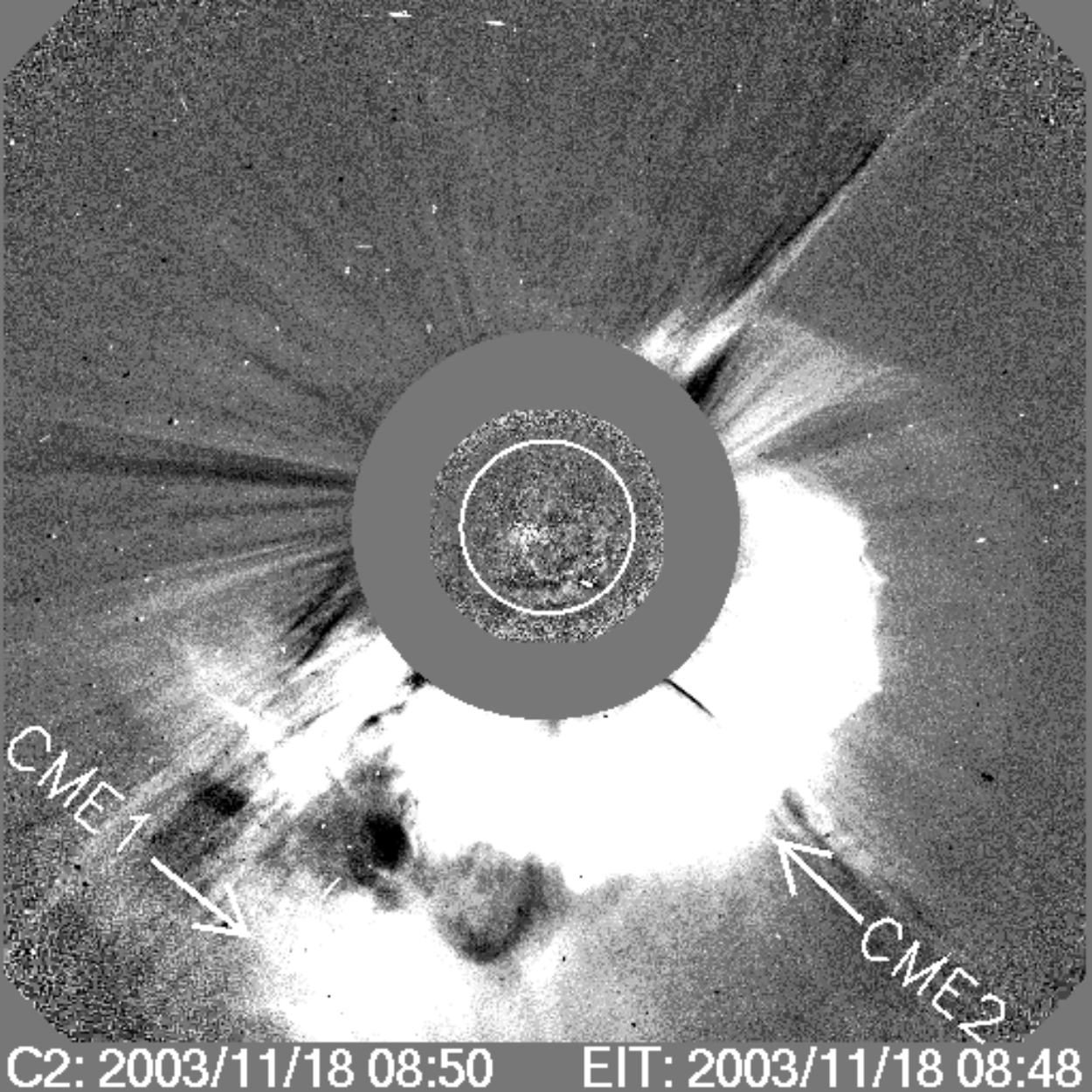}

\includegraphics[width=0.45\textwidth]{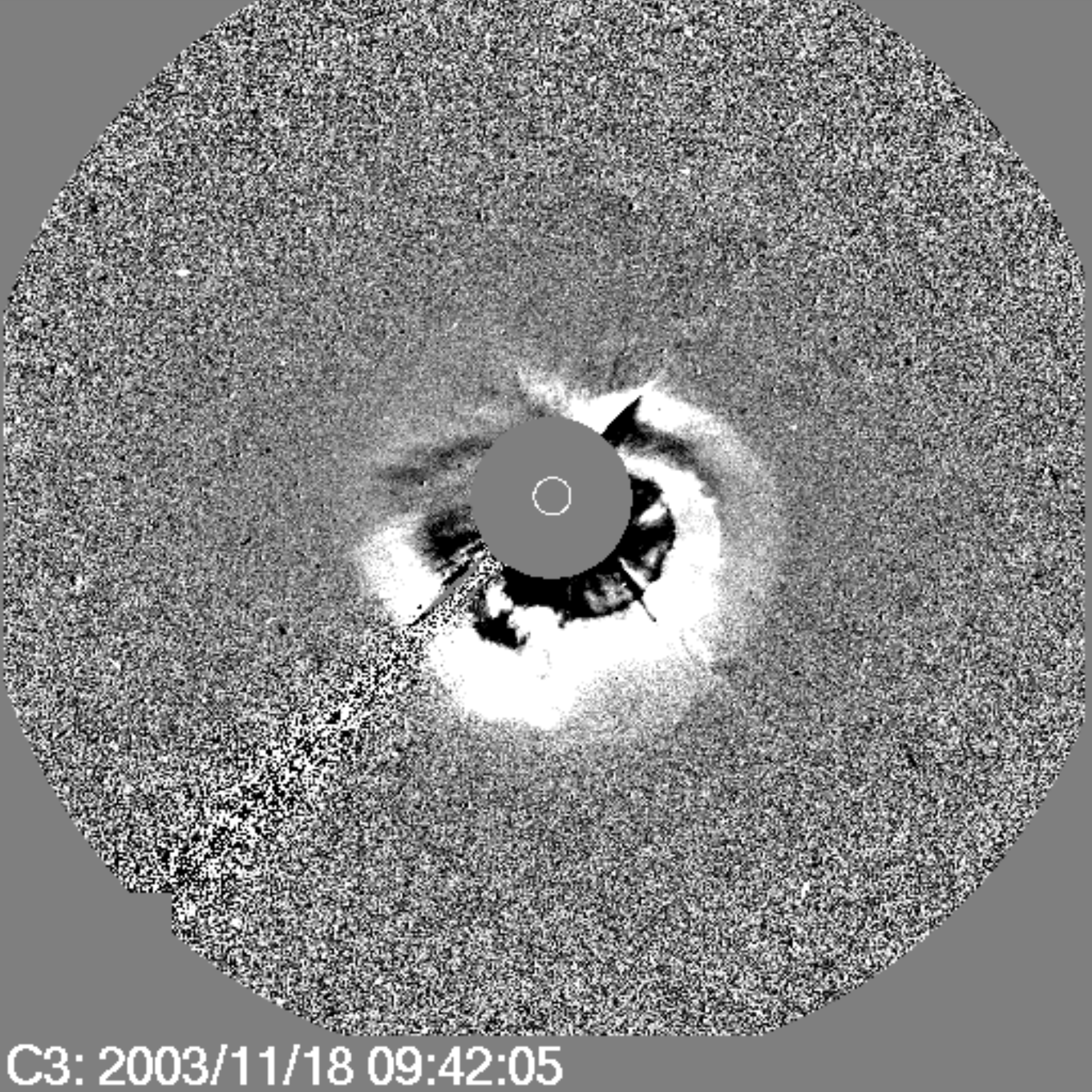}
\includegraphics[width=0.45\textwidth]{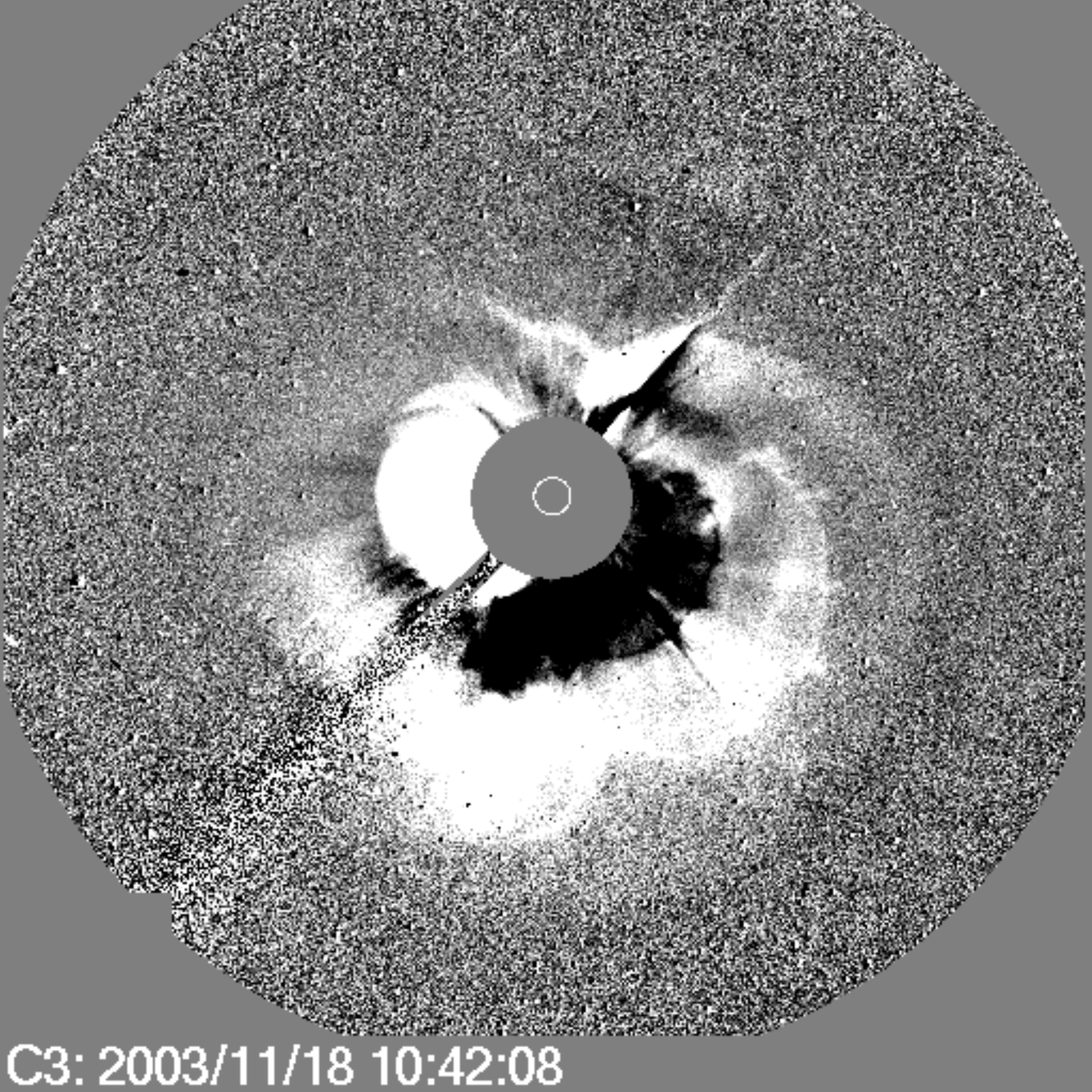}

\caption{Difference images of LASCO C2 and C3 coronagraphs on SOHO showing two CMEs associated with solar flares.}
\label{lasco}
\end{figure}

The Large Angle and Spectrometric Coronagraph (LASCO) observed two CMEs corresponding to the M3.2 and M3.9 flares. Figure \ref{lasco} shows the sequences of LASCO C2 and C3 images. However, the C3.8 flare only produced a rather narrow CME, which was observed at 05:26 UT, having a speed of $\approx$267 km s$^{-1}$. The first partial halo CME onset was observed in the south-east (SE) direction at 08:06 UT in the C2 field of view, in association with the M3.2 flare. The linear speed and acceleration of the CME were respectively 1223 km s$^{-1}$ and 37.8 m s$^{-2}$ . The second full halo CME was observed in the south-west direction at 08:50 UT, corresponding to the M3.9 flare event. It showed a two-part structure, {\it i.e.}, outer and inner parts. However, the outer part may be associated with the shock at the front of the CME. The speed and acceleration of the inner part were 1660 km s$^{-1}$ and --3.3 m s$^{-2}$, respectively. It may be noted that there was another CME in  progress at the eastern limb at 09:50 UT. Since it was very fast (linear speed $\approx$1824 km s$^{-1}$) toward east, it would not have encountered the Earth.

\begin{figure}
\centering
\includegraphics[width=0.9\textwidth]{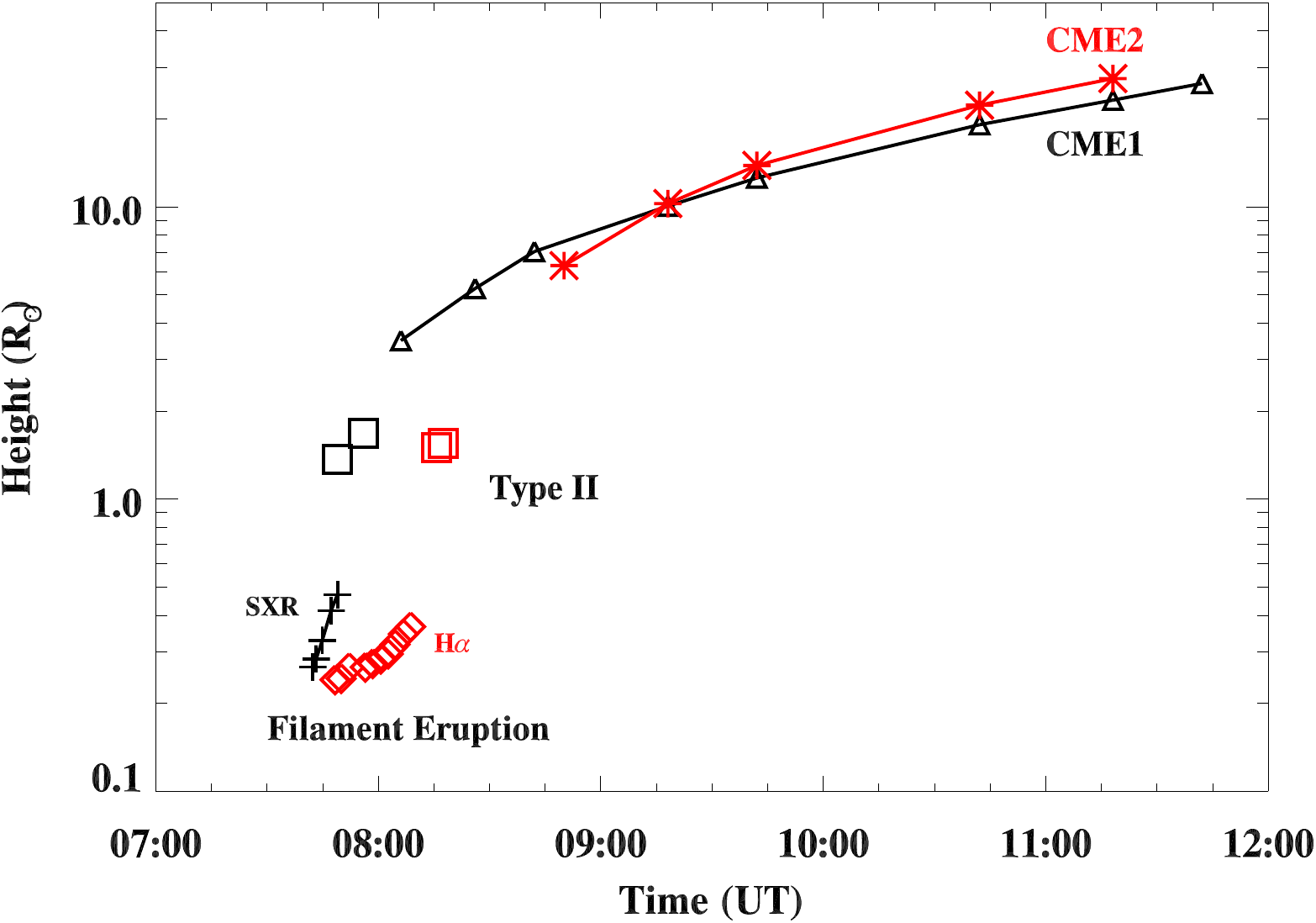}
\caption{The height-time profile of filament eruptions observed in SXI (left) and H$\alpha$ (right, red), two type II radio bursts and CMEs associated with them (LASCO C2 and C3).}
\label{ht_lasco}
\end{figure}

Figure \ref{ht_lasco} displays the height-time profiles for CME1 and CME2. These CMEs interacted at $\approx$09:20 UT at a height of $\approx$10 R$_\odot$. The interaction was clear in the LASCO C3 field of view. In order to see the association of these CMEs with the filament eruptions, we also plotted the data on filament eruptions (one observed in the soft X-rays and another from  H$\alpha$). The  height-time profiles of CMEs show association with the extrapolation of filament eruption in space and time. However, the {\it Wind}/WAVES dynamic spectrum shows the enhanced radio signatures of CME-CME interaction at $\approx$09:00 UT in the interplanetary medium, consistent with LASCO observations. The radio signatures are likely to be associated with the sheath in between the shock and CME as well as CMEs \cite{gopals2001}.
\begin{figure}
\centering
\includegraphics[width=0.9\textwidth]{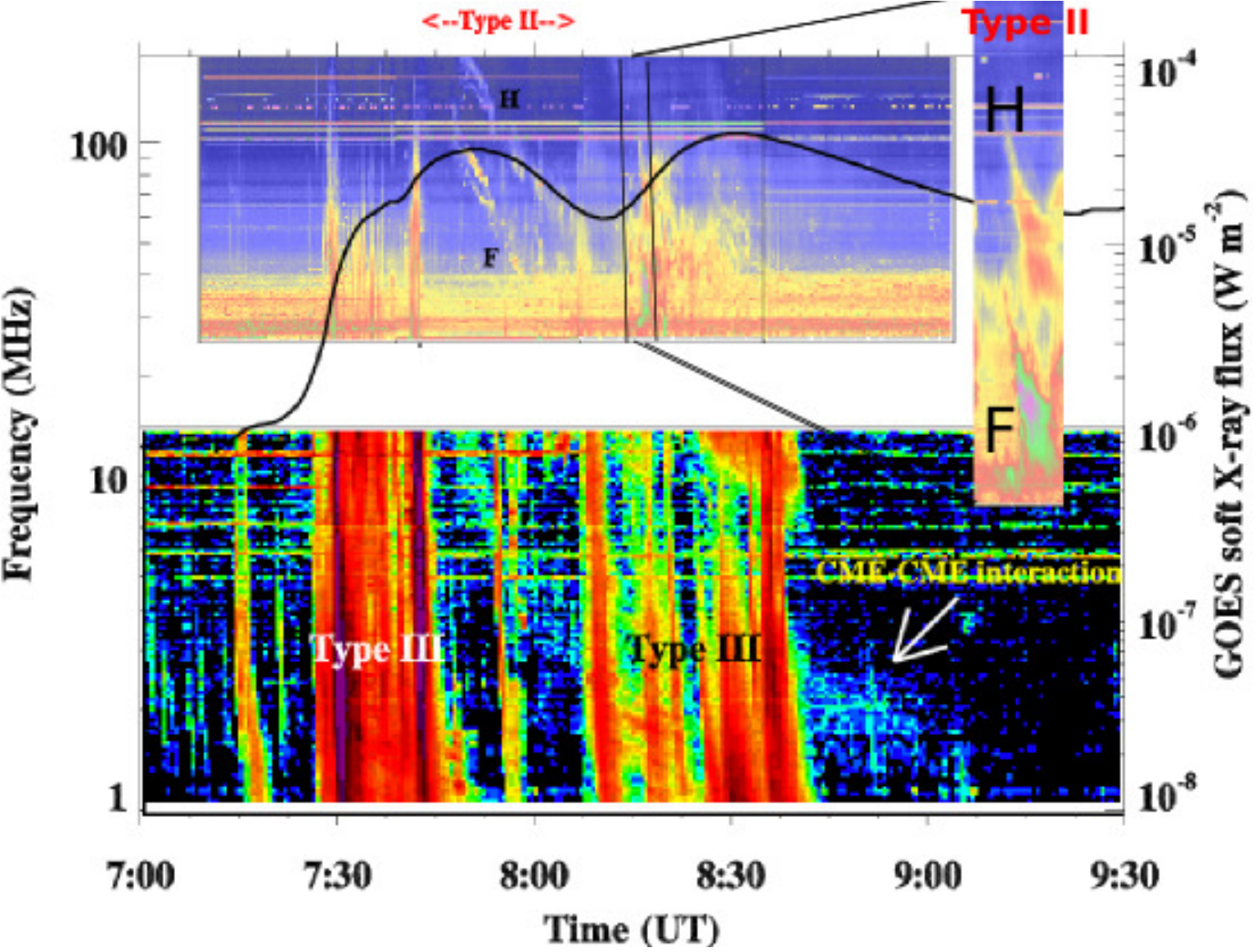}
\caption{Composite radio spectrum from {\it Wind}/WAVES (1-14 MHz) and Learmonth (25-180 MHz) plotted against GOES soft X-ray flux profile (1-8 \AA) on 18 November 2003. This plot shows the presence of two type II bursts during both ejections of the filament. The enlarged view of the second type II burst is shown on the right-hand side.}
\label{spectra}
\end{figure}


\section{Radio Signatures of the Event}
 We use the radio spectral data from Learmonth (25-180 MHz) observatory to study the coronal signatures of the events. Figure \ref{spectra} displays the composite radio spectra from  Learmonth (25-180) and {\it Wind}/WAVES (1-14 MHz) plotted against GOES soft X-ray flux on 18 November 2003.
 It shows two groups of type III radio bursts in the periods of 07:14--07:16 and 07:25--07:44 UT, respectively, two type II bursts during 07:47--07:59 UT and 08:15 UT, and a type IV radio burst during 08:10 UT and 08:30 UT. For the type II radio bursts, the dynamic spectrum shows that the fundamental emission was much stronger than the second harmonic. It may be noted that both fundamental and second harmonic showed band splitting. The type II band splitting appears if the plasma emission occurs from both upstream \cite{smith1971} and downstream \cite{tidman1965,tidman1966} of the coronal shock front, which is used as a proxy to measure the coronal magnetic field strength (e.g., \opencite{cho2007} and references cited therein). We have used the Learmonth solar radio spectral data to measure the shock speed using the frequency drift in both type II band structure alongwith the Newkirk coronal electron density model \cite{newkirk1970}. The estimated shock speeds in both bursts (using the drift-rate of fundamental band structure) were $\approx$400 and $\approx$2200 km s$^{-1}$, respectively.  The observed type III burst at 07:30 UT is basically due to acceleration of electrons along the open  field lines during the reconnection at the eastern leg of the filament. After that event the eastern part of the filament detached and it moved away from the active region. The speed of the first ejection was $\approx$330--370 km s$^{-1}$ measured from TRACE and SXI, respectively. The first coronal type II burst observed 07:47--07:59 UT, indicates the formation of shock in the corona at the front of the filament. It may be noted that the second type II (shown in the right-hand side of Figure \ref{spectra}) was generated by a very high-speed shock. However, the second ejection showed a low speed ($\approx$100 km s$^{-1}$) in the low corona, but generated a very high-speed CME ($\approx$1660 km $^{-1}$). On the other hand, the first ejection ($\approx$330-370 km s$^{-1}$) generated a low-speed shock ($\approx$400 km s$^{-1}$) in the corona and high speed CME ($\approx$1223 km s$^{-1}$). This difference shows that the magnetic energy associated with the second ejection may be larger in comparison to the first one.  We estimate the coronal heights for both shocks using the Newkirk density model \cite{sil2007} and plotted them with time. Figure \ref{ht_lasco} shows a close association of both type II bursts with the two ejections (eastern and central parts) of the filament. In addition, a coronal wave (CW) with complex disturbances was observed during 08:00-08:48 UT propagating from the eruption center southward in the EIT 195 \AA \ and was also detectable in the 304 \AA \ SPIRIT band, but much weaker \cite{grechnev2005}.       

In order to investigate any interplanetary signature of this dynamic event, we use the radio dynamic spectrum from space-based {\it Wind}/WAVES in the frequency range 1--14 MHz, which probes the plasma between $\approx$2-10$R_\odot$. It shows a bunch of type III bursts during both flares. The radio flux enhancement at around 09:00 UT may be due to the interaction of two CMEs at $10 R_\odot$ ({\it e.g.}, \opencite{gopals2001}).  Therefore, this section provides a brief overview of coronal and interplanetary radio bursts associated with the flares and CMEs. It is interesting that both ejections from the same filament showed good association with two coronal type II radio bursts. This is a good example of piston-driven shocks in the corona associated with both ejections. The speed of the ejected filaments (speed $\approx$350 and $\approx$100 km s$^{-1}$) is sufficient enough to generate a shock wave at about $1 R_\odot$. Moreover, the numerical simulation by \inlinecite{chen2005} showed that a flux rope at a speed of 100 km s$^{-1}$ generates a shock wave, which propagates faster than the filament speed.

\begin{figure}
\centering
\includegraphics[width=0.8\textwidth]{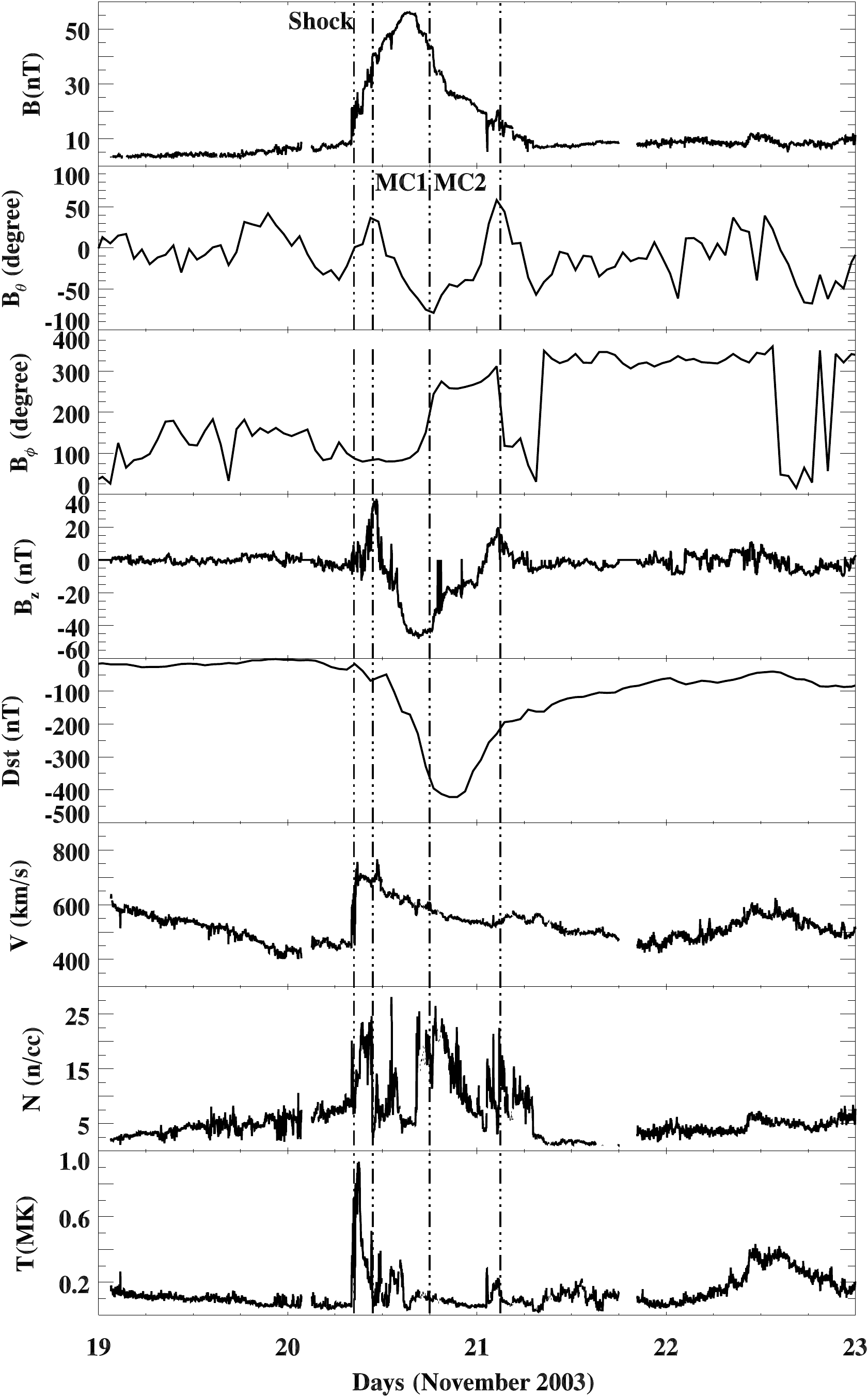}
\caption{The interplanetary observations of magnetic field strength $B$, components $B_\theta$, $B_\phi$, $Dst$ index, solar wind speed, proton density and temperature during 19--23 November 2003.}
\label{omni}
\end{figure}

\section{Interplanetary Consequences}
\begin{figure}
\centering
\includegraphics[width=0.51\textwidth]{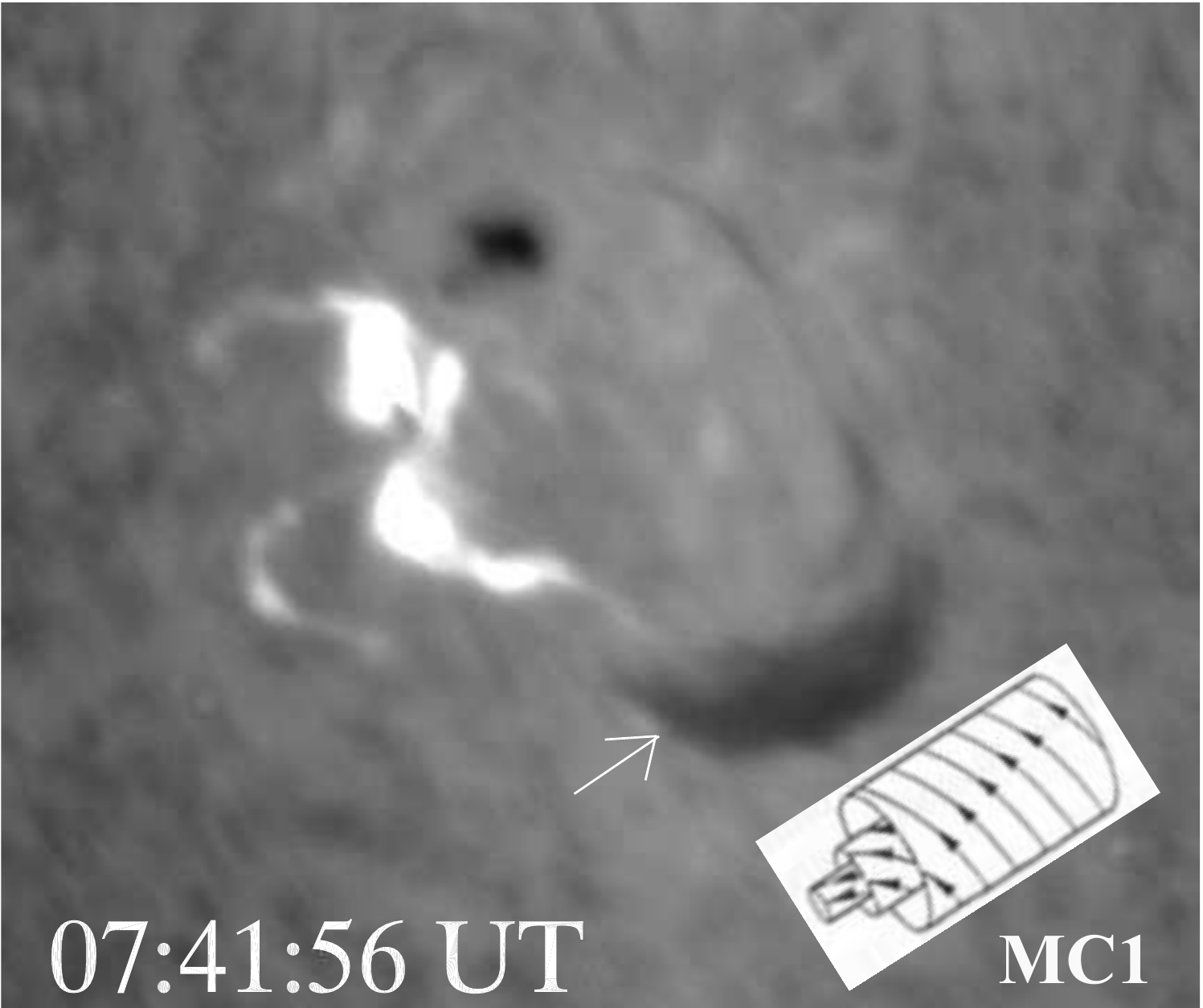}
\includegraphics[width=0.355\textwidth]{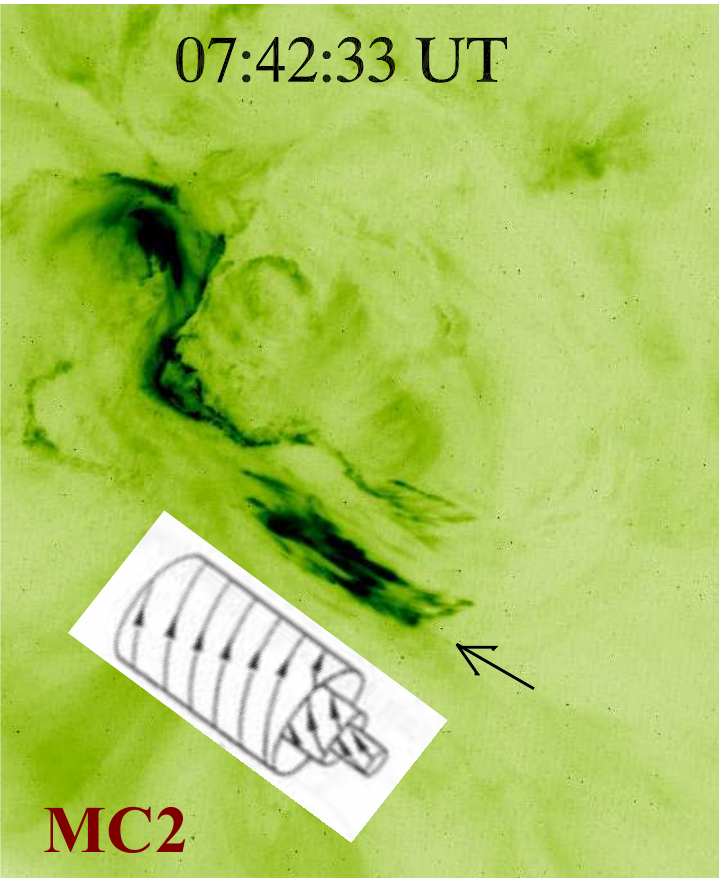}
\caption{The linkage of two merged magnetic clouds in the interplanetary medium with the solar sources.}
\label{linkage}
\end{figure}

\begin{figure}
\centering
\includegraphics[trim=1mm 1mm 1mm 1mm, clip,width=1.0\textwidth]{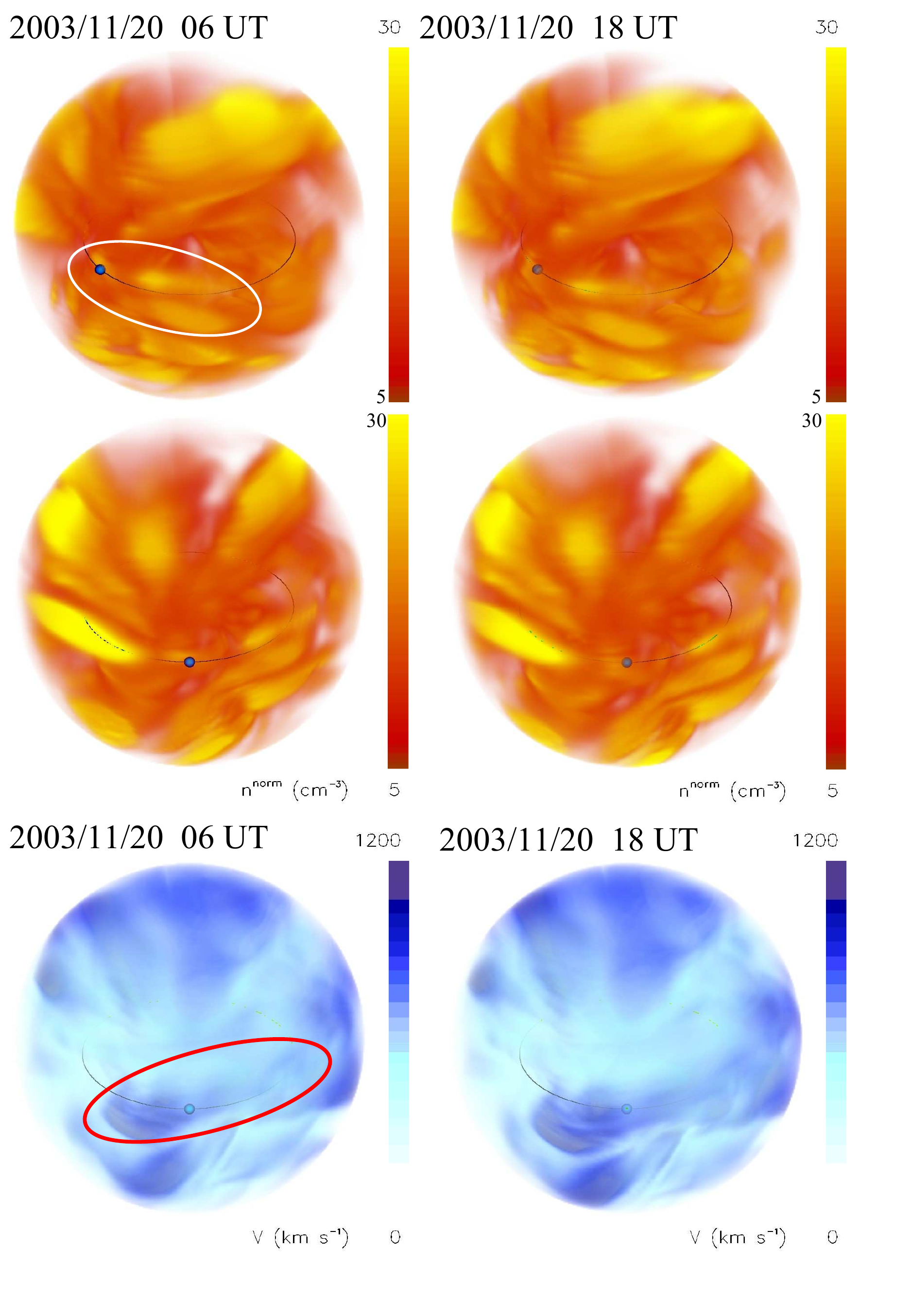}

\caption{ Three-dimensional density (top and middle panels) and velocity (bottom panel) images of the solar wind reconstructed from Ooty IPS observations, showing the merged structure of two CMEs (shown inside the ellipse) hitting the Earth on 
20 November 2003, at around 6 UT.}
\label{ips}
\end{figure}

The interplanetary signatures of the above events have been obtained 
 from {\it in-situ} spacecraft measurements at 1 AU. Figure \ref{omni} shows 
the interplanetary parameters of the CME and the associated shock. The
 arrival of the shock can be seen by the sudden increase in the solar wind
 speed, density, and magnetic field and it is marked by a vertical line at 07:27 UT 
 on 20 November 2003. The azimuthal component of the magnetic field
 ($B_\theta$) showed two smooth rotations, which reveal the presence of 
 two merging magnetic clouds at 1 AU (indicated by MC1 and MC2).  The first magnetic cloud rotated smoothly from south to north with $\phi$=90$^\circ$. Therefore, the cloud was of NES type with right-handed (positive) helicity. The second cloud rotated smoothly from south to north with $\phi$=270$^\circ$. So, this cloud was of SWN type with right-handed (positive) helicity \cite{bothmer1997}. To link these magnetic clouds with their solar sources, we draw schematically the  two clouds over the H$\alpha$ and TRACE 195 \AA \ images respectively. The second ejection from the central part of the filament confirms the linkage with the first magnetic cloud as it contained the big flux rope structure moving with high speed. First ejection, {\it i.e.} the first part of the filament, was linked with the second magnetic cloud (refer to Figure \ref{linkage}).  The scenario shown by the {\it in-situ}
 measurements is consistent with the 3-D images obtained from the
 interplanetary scintillation (IPS) measurements made with the Ooty
 Radio Telescope (ORT), operating at 327 MHz \cite{mano2001,mano2006,mano2010}.
The sequence of IPS images from $\approx50 R_\odot$ to a distance beyond the
orbit of the Earth reveals the propagation of structures associated with
the merged CMEs in the interplanetary medium. Figure \ref{ips} displays 
the reconstructed 3-D distributions of solar wind density and speed within a sphere of 3-AU diameter at two time intervals, {\it i.e.}, 6 and 18 UT on 20 
November 2003. These images provide a remote view at an angle of $\approx$30$^\circ$ 
above the ecliptic plane. The orbit of the Earth is shown by a dark circle, on 
which the location of the Earth is indicated by a filled blue circle. In this figure, 
the top and middle rows show the normalized solar wind density distribution and 
the bottom row displays the corresponding solar wind speed on 20 November 2003. 
Two images at the top are for a view of $\approx$45$^\circ$ west of the 
Sun-Earth line, whereas images in the middle and bottom rows show the 
heliospheric view along the Sun-Earth line. The above images evidently show 
the structures with enhanced density and speed associated with the interacted CMEs 
in the interplanetary space. The arrival of the CME structures inferred from 
the IPS observations and their average speed are in agreement, respectively, 
with the onset of the magnetic cloud and its speed from the {\it in-situ} measurements 
at 1 AU (refer to Figure 14). In Figure 16, the intense 
density structures seen above and below the ecliptic plane (top and middle rows)
are due to the large-amplitude warping of the current sheet and its high density 
and low speed streams (refer to the bottom row).

The polar angle of the interplanetary magnetic field obtained from
 the {\it in situ} data indicates that the cloud was inclined to the ecliptic plane 
 ($\approx$74$^\circ$) and pointed persistently southward as the cloud passed 
the spacecraft. In addition, the azimuthal angle showed a smooth rotation 
 from east to west, with a step at the merging point. The position of the 
 flux rope with respect to the ecliptic is also shown by the IPS images. 
 The shock seems to be associated with the merged magnetic clouds. As the
 cloud crossed the Earth orbit, the solar wind speed changed from 700 to 500 km s$^{-1}$,
 which shows a slow expansion of the cloud. The estimated shock speed is 
 $\approx$800 km s$^{-1}$. The merged magnetic cloud has caused an intense 
geomagnetic storm. The ring current index ($Dst$ index) went down to 
$-472$ nT. The maximum field intensity is about 60 nT, and the field mostly
 pointed to the south ($B_z\approx-53$ nT), which has caused the intense 
 storm. It has remained southward for about 13.5 h. The reasons for the
 intense storms are (i) merged clouds, (ii) their combined speed, (iii) intense 
 shock, and (iv) prolonged as well as (v) intense southward-pointed field.

\section {Discussion and Conclusions}

We have presented results of H$\alpha$, TRACE, and SXI measurements of two consecutive flares (M3.2 and M3.9) on 18 November 2003. The comparatively higher intensity in H$\alpha$ emission of the M3.2 flare than the M3.9 event suggests the reconnection processes in the lower atmosphere, followed by the ejection of the flux rope above the two-ribbon corridor at the flare site. In the LASCO field of view, the CME propagated at $\approx$1220 km s$^{-1}$. The eastern part of the filament seen in soft X-rays (SXI) and EUV (TRACE) showed, respectively, speeds of 330 and 370 km s$^{-1}$, which confirm a heavy acceleration at the time of CME onset. The above ejection took place at the eastern leg of the filament, where the twist was high as shown by H$\alpha$ data. In H$\alpha$, after the above partial eruption, the filament showed motion away from the AR site, with a speed of $\approx$100 km s$^{-1}$ during 07:48--08:10 UT. It might have caused destabilization in the overlying pre-existing complex coronal field structures and produced the next M3.9 flare of maximum intensity at 08:30 UT. The full halo CME in association with this event propagated faster than the previous one at 1660 km s$^{-1}$. However, its propagation towards the south-east direction and the nearly east-west orientation of the associated flare ribbons indicate that (i) the first CME originated at the leg of the filament and (ii) the second CME was associated with the flux rope lying above the ribbons of the flare. In these events both tether cutting and emergence of flux beneath the filament system have played important roles. 

The interaction between these two flux ropes and their merging have led to a rather big magnetic cloud. Therefore we were able to detect close linkage between the cloud and the solar eruptions taking place during the flare events. The magnetic cloud properties observed at 1 AU suggest that the central part of the filament produced the first magnetic cloud (NES) whereas the eastern part of the filament produced the second magnetic cloud (SWN), both having positive helicity.  \inlinecite{wang2006} also compared the tilt angle of the curved filament with the tilt angle of the first magnetic cloud and found that the orientation of the magnetic cloud was almost parallel to the central part of the filament. Our study is also in good agreement with their study. Moreover, \inlinecite{chandra2010} estimated the helicity of the active region which was negative and put a question mark on the positive helicity of the magnetic cloud. However, they estimated the localized helicity of the active region and found the positive helicity ejection at the southern part of the active region {\it i.e.} the source region of flares. The first cloud had a more favorable magnetic topology to initiate the magnetic reconnection with Earth's magnetic field, whereas the second cloud could help to sustain the reconnection for a longer time. Therefore, the prolonged reconnection of the resultant magnetic cloud with Earth's magnetosphere has caused a very intense storm ($Dst_{\rm peak}=-472$ nT) of the solar cycle 23. 

This study illustrates the need of combining solar as well as {\it in-situ} interplanetary propagation of the ejecta for the understanding of the physical processes. \inlinecite{ermol2005} pointed out that the strongest disturbance observed in the interplanetary medium was apparently caused by the interaction between two or three magnetic clouds produced by the series of flares and CMEs on 18 November, 2003. \inlinecite{gopal2005} suggested that the largest geomagnetic storm of solar cycle 23 was caused by a fast
and wide CME ($\approx$1660 km s$^{-1}$) originating from a region close to the disk center, which resulted in a highly inclined MC
with its axial field pointing almost always southward.
  Previous studies of the same event,
 but mostly limited, have concluded the presence of only one-
ejecta ({\it e.g}, \opencite{gopal2005}, \opencite{zhang2007}, \opencite{echer2008b}, \opencite{chandra2010}). However, here we confirm the two merged magnetic clouds associated with two ejecta from the Sun. Needless to say, more detailed study of such events should be carried out with recent space observations ({\it e.g.}, {\it Hinode}, STEREO, and SDO) to shed more light on the eruption process and their interplanetary consequences. Finally, this event provides a unique opportunity to study the flare-CME relationship, the interaction as well as merging of solar wind magnetic structures which are important in understanding the space weather. 


\begin{acks}
We are grateful to the referee for helpful and constructive
comments. The authors thank the observing and engineering staff of Radio Astronomy Centre in making the IPS observations. We also thank B. Jackson and the UCSD team for the IPS tomography analysis package. SOHO (EIT, LASCO, and MDI images) is a project of international cooperation between ESA and NASA. PKM acknowledges the partial support for this study by CAWSES-India Program, which is sponsored by ISRO. We thank CEFIPRA Project 3704-1 for its support to this study on ``Transient events in Sun Earth System" during our bilateral collaboration. We also acknowledge {\it Wind}/WAVES and Learmonth observatory team for providing radio dynamic spectra.
\end{acks}

\bibliographystyle{spr-mp-sola}
\bibliography{reference}  
\end{article}
\end{document}